\documentclass[aps,prd,showpacs,twocolumn,floatfix,nofootinbib]{revtex4-1}
\usepackage[dvipdfmx]{graphicx}
\usepackage{amsmath,amssymb,siunitx}
\usepackage{color,ulem}
\usepackage{graphicx}
\usepackage{bm,latexsym,amsmath,amssymb,amsfonts,mathrsfs}
\usepackage{dcolumn}   
\usepackage{color}

\newcommand{\simgt}{\lower.5ex\hbox{$\; \buildrel > \over \sim \;$}}
\newcommand{\simlt}{\lower.5ex\hbox{$\; \buildrel < \over \sim \;$}}
\newcommand{\solM}{M_{\odot}}

\begin{document}
\title{Multipole tidal effects in the post-Newtonian gravitational-wave phase of compact binary coalescences}
\author{Tatsuya Narikawa}
\email{narikawa@icrr.u-tokyo.ac.jp}
\affiliation{
$^1$Institute for Cosmic Ray Research, The University of Tokyo, Chiba 277-8582, Japan\\
}
\date{\today}

\begin{abstract}
We present the multipole component form of the gravitational-wave tidal phase for compact binary coalescences (MultipoleTidal), 
which consists of the mass quadrupole, the current quadrupole, and the mass octupole moments.
We demonstrate the phase evolution and the phase difference between the tidal multipole moments (MultipoleTidal) and the mass quadrupole (PNTidal) as well as the numerical-relativity calibrated model (NRTidalv2).
We find the MultipoleTidal gives a larger phase shift than the PNTidal, and is closer to the NRTidalv2.
We compute the matches between waveform models to see the impact of the tidal multipole  moments on the gravitational wave phases. 
We find the MultipoleTidal gives larger matches to the NRTidalv2 than the PNTidal, in particular, for high masses and large tidal deformabilities.
We also apply the MultipoleTidal model to binary neutron star coalescence events GW170817 and GW190425.
We find that the current quadrupole and the mass octupole moments give no significant impact on the inferred tidal deformability.
\end{abstract}


\maketitle

\section{Introduction}
\label{sec:Introduction}
The post-Newtonian (PN) tidal waveform of compact binary coalescences for the mass quadrupole 
have been derived up to 2.5PN (relative 7.5PN) order for phase~\cite{Damour:2012yf, Bini:2012gu, Agathos:2015uaa, Henry:2020ski} and have used in analyses of binary neutron star (BNS) coalescence signals detected by the Advanced LIGO and Advanced Virgo detectors, GW170817 and/or GW190425~\cite{TheLIGOScientific:2017qsa, Abbott:2018wiz, De:2018uhw, LIGOScientific:2018mvr, Abbott:2020uma, LIGOScientific:2020ibl, Dai:2018dca, Narikawa:2018yzt, Narikawa:2019xng, Gamba:2020wgg, Breschi:2021wzr,Narikawa:2022saj} (see also Refs.~\cite{LIGOScientific:2018ehx, Pratten:2019sed, Pan:2020tht,Pradhan:2022rxs}).

The complete and correct PN tidal phase up to relative 7.5PN order for the mass quadrupole, the current quadrupole, and the mass octupole interactions have been derived using the PN-matched multiplolar-post-Minkowskian formalism~\cite{Henry:2020ski}.
In our previous study~\cite{Narikawa:2021pak}, 
we rewrote the complete and correct HFB (the abbreviation for Henry, Faye, and Blanchet) form to the quadrupole component form
for the mass quadrupole interactions (hereafter PNTidal) as a function of the dimensionless tidal deformability for the individual stars, $\Lambda$, in a convenient way for data analyses.
And we have reanalyzed the data around low-mass events identified as binary black holes  by using the corrected version of the PNTidal model
to test the exotic compact object hypothesis.
Here, uncalculated coefficients at 5+2PN order are completed and coefficients at 5+2.5PN order are corrected \cite{Henry:2020ski, Narikawa:2022saj, Narikawa:2021pak}.
The PNTidal model has also been used in analyses of BNS signals \cite{Narikawa:2022saj} (see also Ref. \cite{Breschi:2021wzr}).
In Ref. \cite{Pradhan:2022rxs}, the impact of the updated equation-of-state (EOS) insensitive relations for multipole tidal deformabilities and for $f$-mode dynamical tidal correction and the mass quadrupole in the context of BNS signals has been studied.

In this study, we present the multipole component form of the tidal phases (hereafter MultipoleTidal), which consists of the mass quadrupole, the current quadrupole, and the mass octupole moments.
By using the multipole component form, we compare the match between the MultipoleTidal model and the NRTidalv2 model, which is the numerical relativity (NR) calibrated model for the tidal part, 
and the match between the PNTidal model and the NRTidalv2 model.
We find that the match between the MultipoleTidal model and the NRTidalv2 model is better than the match between the PNTidal model and the NRTidalv2 model, in particular, for high mass and large tidal deformabilities.
We also apply the MultipoleTidal model to BNS coalescence events GW170817 and GW190425 to investigate the impact of the model on the parameter estimation of the current events.

The outline of the paper is as follows.
In Sec.~\ref{sec:WF}, we present the multipole component form of the tidal phases, MultipoleTidal, which consists of the mass quadrupole, the current quadrupole, and the mass octupole moments.
In Sec.~\ref{sec:Comparison}, we demonstrate the impact of the multipole tidal contributions by comparing the MultipoleTidal model with the PNTidal model and the NRTidalv2 model.
In Sec.~\ref{subsec:ComparisonPhase}, we present the phase evolution for the MultipoleTidal model, the PNTidal model, and several NR calibrated tidal models, and the phase difference between the MultipleTidal and the PNTidal. 
In Sec.~\ref{subsec:Match}, we compare the match between the MultipoleTidal and the NRTidalv2 and the match between the PNTidal and the NRTidalv2. 
In Sec.~\ref{subsec:GW170817}, we show the parameter estimation results of BNS coalescence events GW170817 and GW190425 with the MultipoleTidal.
Sec.~\ref{sec:summary} is devoted to a conclusion.
In Appendix \ref{sec:EOS-insensitive-relations}, we summarize the EOS-insensitive relations for the tidal multipole moments.
In Appendix \ref{sec:WFsys}, we present two-dimensional posterior distributions of source parameters for GW170817 and GW190425 obtained with three different tidal waveform models: MultipoleTidal, PNTidal, and NRTidalv2.


\vspace{2pt}
\section{Waveform models for inspiraling binary neutron stars}
\label{sec:WF}
In this section, we show multipole tidal effects in the GW phases .
First, we briefly introduce the HFB form derived in Ref. \cite{Henry:2020ski}.
Then, we present the multipole component form.
Finally, we present the identical-NS form, which focus on dominant contributions and is useful to fit NR waveforms.

\subsection{Multipole tidal interactions}
\label{sec:tidal}
The tidal polarizability coefficients are denoted as \cite{Henry:2020ski}
\begin{eqnarray}
 G\mu_A^{(2)} \equiv \left( \frac{G m_{A}}{c^2} \right)^5 \Lambda_A = \frac{2}{3} k_A^{(2)} R_A^5,
\end{eqnarray}
for the mass quadrupole moment,
\begin{eqnarray}
 G\sigma_A^{(2)} \equiv \left( \frac{G m_{A}}{c^2} \right)^5 \Sigma_A = \frac{1}{48} j_A^{(2)} R_A^5, 
\end{eqnarray}
for the current quadrupole moment, and
\begin{eqnarray}
 G\mu_A^{(3)} \equiv \left( \frac{G m_{A}}{c^2} \right)^7 \Lambda_A^{(3)} = \frac{2}{15} k_A^{(3)} R_A^7,
\end{eqnarray}
for the mass octupole moment, where $k_A^{(2)}$, $j_A^{(2)}$, and $k_A^{(3)}$ are Love numbers 
and $\Lambda_A$, $\Sigma_A$, and $\Lambda_A^{(3)}$ are dimensionless multipole tidal deformability parameters, of each object with a component mass $m_{A}$ and a radius $R_A$.

\subsection{HFB form}
\label{sec:HFB-form}
The complete and correct form up to 2.5PN (relative 7.5PN) order
for the mass quadrupole, the current quadrupole, and the mass octupole contributions to the GW tidal phases have been derived \cite{Henry:2020ski}. 
In this subsection, we briefly introduce the HFB form.

The polarizability parameters are redefined for convenience as 
\begin{eqnarray}
 \mu_\pm^{(l)} = \frac{1}{2} \left( \frac{m_B}{m_A} \mu_A^{(l)} \pm \frac{m_A}{m_B} \mu_B^{(l)} \right), 
\label{eq:mu_pm}
\end{eqnarray}
for the mass multipole moments, and 
\begin{eqnarray}
 \sigma_\pm^{(l)} = \frac{1}{2} \left( \frac{m_B}{m_A} \sigma_A^{(l)} \pm \frac{m_A}{m_B} \sigma_B^{(l)} \right),
\label{eq:sigma_pm}
\end{eqnarray}
for the current multipole moments, where $l$ is a positive integer. 
For two identical NS (NS is the abbreviation of a neutron star), $\mu_+^{(l)} = \mu_A^{(l)} = \mu_B^{(l)}$ and $\mu_-^{(l)} = 0$ and
$\sigma_+^{(l)} = \sigma_A^{(l)} = \sigma_B^{(l)}$ and $\sigma_-^{(l)} = 0$.
The adimensionalized parameters corresponding to Eqs. (\ref{eq:mu_pm}) and (\ref{eq:sigma_pm}) are defined as 
\begin{eqnarray}
 \tilde{\mu}_\pm^{(l)} = \left( \frac{c^2}{G M} \right)^{2l+1} G \mu_\pm^{(l)}, 
 \label{eq:tilde_mu}
\end{eqnarray}
and
\begin{eqnarray}
 \tilde{\sigma}_\pm^{(l)} = \left( \frac{c^2}{G M} \right)^{2l+1} G \sigma_\pm^{(l)},
 \label{eq:tilde_sigma}
\end{eqnarray}
respectively, where $M = m_A + m_B$ is the total mass.

The HFB-form for the tidal phases is derived as 
\begin{widetext}
\begin{eqnarray}
&& \Psi_\mathrm{HFB}(f) = \nonumber \\
 && - \frac{9}{16 \eta^2} x^{5/2} \left\{ (1 + 22\eta) \tilde{\mu}_+^{(2)} + \Delta \tilde{\mu}_-^{(2)}\right.\nonumber \\
 && + \left[ \left( \frac{195}{112} + \frac{1595}{28} \eta +\frac{325}{84}\eta^2 \right) \tilde{\mu}_+^{(2)} 
 + \Delta \left( \frac{195}{112} + \frac{4415}{336}\eta \right) \tilde{\mu}_-^{(2)} 
 + \left( -\frac{5}{126} + \frac{1730}{21}\eta \right) \tilde{\sigma}_+^{(2)} - \frac{5}{126}\Delta \tilde{\sigma}_-^{(2)} \right] x \nonumber \\
 && - \pi \left[ (1 + 22\eta) \tilde{\mu}_+^{(2)} + \Delta \tilde{\mu}_-^{(2)} \right] x^{3/2} \nonumber \\
 && + \left[ \left( \frac{136190135}{27433728} + \frac{978554825}{4572288}\eta - \frac{281935}{6048}\eta^2 +\frac{5}{3}\eta^3 \right) \tilde{\mu}_+^{(2)} + \Delta \left( \frac{136190135}{27433728} + \frac{213905}{2592}\eta + \frac{1585}{1296}\eta^2 \right) \tilde{\mu}_-^{(2)} \right. \nonumber \\
 && \left. + \left( - \frac{745}{4536} + \frac{1933490}{5103}\eta - \frac{3770}{81}\eta^2 \right) \tilde{\sigma}_+^{(2)} 
 + \Delta \left( - \frac{745}{4536} + \frac{19355}{243}\eta \right) \tilde{\sigma}_-^{(2)} + \frac{1000}{27}\eta \tilde{\mu}_+^{(3)} \right] x^2 \nonumber \\
&& \left. + \pi \left[ \left( - \frac{397}{112}- \frac{5343}{56}\eta + \frac{1315}{42}\eta^2 \right) \tilde{\mu}_+^{(2)} + \Delta \left( - \frac{397}{112} - \frac{6721}{336}\eta \right) \tilde{\mu}_-^{(2)} + \left( \frac{2}{21} - \frac{8312}{63}\eta \right) \tilde{\sigma}_+^{(2)} + \frac{2}{21} \Delta \tilde{\sigma}_-^{(2)} \right] x^{5/2} \right\}, \nonumber \\
\label{eq:HFB}
\end{eqnarray}
\end{widetext}
where 
$x=[\pi G M (1+z) f / c^3]^{2/3}$ is the dimensionless PN expansion parameter, 
$z$ is the source redshift,
$\eta=m_1 m_2 / (m_1+m_2)^2$ is the symmetric mass ratio, and $\Delta = (m_A - m_B) / M$ is the normalized mass difference~\footnote{Following Ref.~\cite{Henry:2020ski_v4}, Eq.~(\ref{eq:HFB}) has been modified in the latest version, which is described in the Erratum~\cite{Narikawa_Erratum}. Two coefficients of 7PN order have been modified.}.

\subsection{Multipole component form of tidal phase}
\label{sec:Component-form}
In our previous study~\cite{Narikawa:2021pak}, 
we rewrote the HFB form to the quadrupole component form for the mass quadrupole interactions
as a function of the dimensionless tidal deformability for the component stars, $\Lambda_{A,B}$, which is called the PNTidal model.
In this study, we extend the PNTidal to the multipole moments, which consists of 
the mass quadrupole, the current quadrupole, and the mass octupole contributions, which is called the MultipoleTidal model.

By using $X_{A,B} = m_{A,B} / M$, the adimensionalized parameters Eqs. (\ref{eq:tilde_mu}) and (\ref{eq:tilde_sigma}) are rewritten as
\begin{eqnarray}
 2\tilde{\mu}_\pm^{(2)} = (1-X_A) X_A^4 \Lambda_A \pm (A \leftrightarrow B), 
\end{eqnarray}
for the mass quadrupole moment,
\begin{eqnarray}
 2\tilde{\sigma}_\pm^{(2)} = (1-X_A) X_A^4 \Sigma_A \pm (A \leftrightarrow B), 
\end{eqnarray}
for the current quadrupole moment, and
\begin{eqnarray}
 2\tilde{\mu}_\pm^{(3)} = (1-X_A) X_A^4 \Lambda_A^{(3)} \pm (A \leftrightarrow B),
\end{eqnarray}
for the mass octupole moment.
Also, $X_B$ and $\Delta$ are rewritten as $X_B=1-X_A$ and $\Delta = 2 X_A - 1$ by using $X_A$.

\subsubsection{Mass quadrupole}
\label{sec:Component-MassQuad}
The leading order effect for the mass quadrupole interactions appears relative 5PN order on GW phases \cite{Hinderer:2007mb,Flanagan:2007ix, Hinderer:2009ca, Vines:2011ud}.
The PN tidal phases for the mass quadrupole have been derived up to 2.5PN (relative 7.5PN) order \cite{Damour:2012yf, Bini:2012gu, Agathos:2015uaa}, and 
completed and corrected up to relative 7.5PN order by Ref. \cite{Henry:2020ski} as shown in Eq. (\ref{eq:HFB}).
The quadrupole component form for the mass quadrupole interactions (PNTidal) is written as a function of $\Lambda_{A,B}$ as~\cite{Narikawa:2021pak,Narikawa:2022saj}
\begin{widetext}
\begin{eqnarray}
 \Psi_\mathrm{Component}^\mathrm{MassQuad}(f) &=& \frac{3}{128\eta} x^{5/2} \Lambda_A X_A^4 
 \left[ -24(12-11 X_A) - \frac{5}{28} (3179-919 X_A - 2286 X_A^2 + 260 X_A^3 ) x \right. \nonumber \\
 && +24 \pi (12 - 11 X_A) x^{3/2} \nonumber \\
 && -5 \left( \frac{193986935}{571536} - \frac{13060861}{381024} X_A - \frac{59203}{378} X_A^2
 - \frac{209495}{1512} X_A^3 + \frac{965}{54} X_A^4 - 4 X_A^5 \right) x^2 \nonumber \\
 &&  \left. + \frac{\pi}{28} (27719 - 22415 X_A + 7598 X_A^2 - 10520 X_A^3) x^{5/2}  \right] + (A \leftrightarrow B).
\label{eq:MassQuad_phase}
\end{eqnarray}
\end{widetext}
%
This is simply written in terms of $\Lambda_{A,B}$, which is measurable via GWs. 
We have implemented it in the waveform section, LALSimulation \cite{Veitch:2014wba}, as part of the LSC Algorithm Library (LAL) \cite{LAL} and used it in data analyses \cite{Narikawa:2021pak,Narikawa:2022saj}\footnote{Similarly to the modification to Eq.~(\ref{eq:HFB}), Eq.~(\ref{eq:MassQuad_phase}) has also been modified in the latest version.}.

\subsubsection{Current quadrupole}
\label{sec:Component-CurrentQuad}
The leading order effect for the current quadrupole interactions appears relative 6PN order on GW phases \cite{Abdelsalhin:2018reg, Banihashemi:2018xfb} (see also Refs. \cite{JimenezForteza:2018rwr,Castro:2022mpw}).
The terms have been completed up to relative 7.5PN order by Ref. \cite{Henry:2020ski} as shown in Eq. (\ref{eq:HFB}).
We rewrite the HFB form to the quadrupole component form 
for the current quadrupole interactions as a function of the dimensionless current quadrupole  parameters for the individual stars, $\Sigma_{A,B}$, as
\begin{widetext}
\begin{eqnarray}
 \Psi_\mathrm{Component}^\mathrm{CurrentQuad}(f) &=& \frac{3}{128\eta} x^{5/2} \Sigma_A X_A^4 
 \left[ - \frac{20}{21} (1037 - 1038 X_A ) x \right. \nonumber \\
 && - \frac{5}{1701} \left( 1220287 - 761308 X_A - 270312 X_A^2 - 190008 X_A^3 \right) x^2 \nonumber \\
 &&  \left. + \frac{16}{21}\pi (2075 - 2078 X_A ) x^{5/2}  \right] + (A \leftrightarrow B). 
\label{eq:CurrentQuad_phase}
\end{eqnarray}
\end{widetext}
As the mass quadrupole, this is simply written in terms of $\Sigma_{A,B}$, which is measurable via GWs.
We implement it in the LALSimulation and use it in our data analyses.

\subsubsection{Mass octupole}
\label{sec:Component-MassOct}
The leading order effect for the mass octupole interactions appears relative 7PN order on GW phases \cite{Abdelsalhin:2018reg, Landry:2018bil}.
The terms have been completed up to relative 7.5PN by Ref. \cite{Henry:2020ski} as shown in Eq. (\ref{eq:HFB}).
We rewrite the HFB form to the octupole component form 
for the mass octupole interactions as a function of the dimensionless mass octupole  parameters for the individual stars, $\Lambda^{(3)}_{A,B}$, as
%
\begin{eqnarray}
&& \Psi_\mathrm{Component}^\mathrm{MassOct}(f) \nonumber \\
&& = \frac{3}{128\eta} x^{5/2} \Lambda_A^{(3)} X_A^4 
 \left[ - \frac{4000}{9} \left( 1 - X_A \right) x^2 \right] \nonumber \\
 && + (A \leftrightarrow B). 
\label{eq:MassOct_phase}
\end{eqnarray}
As the mass quadrupole and the current quadrupole, this is simply written in terms of $\Lambda_{A,B}^{(3)}$, which is measurable via GWs.
We implement it in the LALSimulation and use it in our data analyses.

\subsection{Identical-NS form}
\label{sec:IdenticalNS-form}
For realistic cases, the tidal contributions to the GW phase are dominated by the symmetric contribution in terms of the multipole tidal polarizability (e.g., \cite{Favata:2013rwa,Wade:2014vqa}).
Motivated by this fact, ignoring the asymmetric contributions, the identical-NS form is obtained.

\subsubsection{Mass quadrupole}
\label{sec:IdenticalNS-MassQuad}
The identical-NS form for the mass quadrupole is obtained from Eq.~(\ref{eq:MassQuad_phase}) by replacing $X_{A,B}$ by $1/2$ and $\Lambda_{A,B}$ by $\tilde{\Lambda}$ as \cite{Narikawa:2022saj}
\begin{eqnarray}
&& \Psi_{\mathrm{Identical-NS}}^\mathrm{MassQuad} (f) = 
 \frac{3}{128\eta} \left( - \frac{39}{2} \tilde{\Lambda} \right) x^{5/2} \nonumber \\
 && \times \left[ 1 + \frac{3115}{1248} x - \pi x^{3/2} + \frac{29323235}{3429216} x^2 - \frac{2137}{546} \pi x^{5/2} \right]. \nonumber \\
\label{eq:MassQuad_phase_identical-NS-form} 
\end{eqnarray}
The NR calibrated tidal waveform models: \texttt{KyotoTidal}~\cite{Kawaguchi:2018gvj}, \texttt{NRTidal}~\cite{Dietrich:2017aum, Dietrich:2018uni}, and \texttt{NRTidalv2}~\cite{Dietrich:2019kaq}, are constructed by extension of this form\footnote{Similarly to the modification to Eqs.~(\ref{eq:HFB}) and (\ref{eq:MassQuad_phase}), Eq. (\ref{eq:MassQuad_phase_identical-NS-form}) has also been modified in the latest version.}.
The binary tidal deformability $\tilde{\Lambda}$ is obtained as follows.
By rewriting the leading terms of Eq.~(\ref{eq:MassQuad_phase}) as  
\begin{eqnarray}
&& \mathrm{(the~leading~term~of~Eq.}~(\ref{eq:MassQuad_phase})) \big/ \left( \frac{3}{128\eta} x^{5/2} \right) \nonumber \\
&& = \Lambda_A X_A^4 \left[ -24(12-11 X_A) \right] + (A \leftrightarrow B) \nonumber \\
&& = \left( -\frac{39}{2} \right) \left( -\frac{39}{2} \right)^{-1} (-24) \left[ (12-11 X_A) X_A^4 \Lambda_A \right. \nonumber \\
&& \left. + (A \leftrightarrow B) \right] \nonumber \\
&& = \left( -\frac{39}{2} \right) \frac{16}{13} \left[ (12-11 X_A) X_A^4 \Lambda_A + (A \leftrightarrow B) \right], \nonumber \\
\label{eq:MassQuad_phase_identical-NS-form_derive}
\end{eqnarray}
and comparing it with Eq.~(\ref{eq:MassQuad_phase_identical-NS-form}), 
it is obtained as 
\begin{eqnarray}
 \tilde{\Lambda} = \frac{16}{13} \left[ 
 \left( 12 - 11X_A \right) X_A^4 \Lambda_A + (A \leftrightarrow B)
\right].
\end{eqnarray}
$\tilde{\Lambda}$ is a mass-weighted linear combination of the mass quadrupole component tidal parameters $\Lambda_{A,B}$.
%

\subsubsection{Current quadrupole}
\label{sec:IdenticalNS-CurrentQuad}
As the mass quadrupole, 
the identical-NS form for the current quadrupole is obtained from Eq.~(\ref{eq:CurrentQuad_phase}) by replacing $X_{A,B}$ by $1/2$ and $\Sigma_{A,B}$ by $\tilde{\Sigma}$ as follows:
\begin{eqnarray}
&& \Psi_{\mathrm{Identical-NS}}^\mathrm{CurrentQuad}(f) = 
 \frac{3}{128\eta} \left( - \frac{185}{3} \tilde{\Sigma} \right) x^{5/2} \nonumber \\
 && \times \left[ x + \frac{93538}{20979} x^2 - \frac{8}{5} \pi x^{5/2} \right]. 
\label{eq:CurrentQuad_phase_identical-NS-form} 
\end{eqnarray}
The binary current quadrupole tidal deformability $\tilde{\Sigma}$ is obtained as follows.
By rewriting the leading terms of Eq.~(\ref{eq:CurrentQuad_phase}) as
\begin{eqnarray}
&& \mathrm{(the~leading~term~of~Eq.}~(\ref{eq:CurrentQuad_phase})) \big/ \left( \frac{3}{128\eta} x^{5/2} \right) \nonumber \\
&& = \Sigma_A X_A^4 \left[ -\frac{20}{21} (1037 - 1038 X_A) \right] + (A \leftrightarrow B) \nonumber \\
&& = \left( - \frac{185}{3} \right) \left( - \frac{185}{3} \right)^{-1} \left(-\frac{20}{21}\right) \left[ (1037-1038 X_A) \right. \nonumber \\
&& \left. \times X_A^4 \Sigma_A + (A \leftrightarrow B) \right] \nonumber \\
&& = - \frac{185}{3} \frac{4}{259} \left[ (1037-1038 X_A) X_A^4 \Sigma_A + (A \leftrightarrow B) \right], \nonumber \\
\label{eq:CurrentQuad_phase_identical-NS-form_derive} 
\end{eqnarray}
and comparing it with Eq.~(\ref{eq:CurrentQuad_phase_identical-NS-form}), it is obtained as 
\begin{eqnarray}
 \tilde{\Sigma} = \frac{4}{259} \left[ 
 \left( 1037 - 1038X_A \right) X_A^4 \Sigma_A + (A \leftrightarrow B)
\right]. \nonumber \\
\end{eqnarray}
$\tilde{\Sigma}$ is a mass-weighted linear combination of the current quadrupole component parameters $\Sigma_{A,B}$\footnote{Our definition of $\tilde{\Sigma}$ is different from Eq.~(6) of \cite{Landry:2018bil} by a factor of 2.}.

\subsubsection{Mass octupole}
\label{sec:IdenticalNS-MassOct}
As the mass quadrupole and the current quadrupole, 
the identical-NS form for the mass octupole is obtained from Eq.~(\ref{eq:MassOct_phase}) by replacing $X_{A,B}$ by $1/2$ and $\Lambda_{A,B}^{(3)}$ by $\tilde{\Lambda}^{(3)}$ as
\begin{eqnarray}
&& \Psi_{\mathrm{Identical-NS}}^\mathrm{MassOct}(f) = 
 \frac{3}{128\eta} \left(- \frac{250}{9} \tilde{\Lambda}^{(3)} \right) x^{5/2} \left[ x^2 \right]. \nonumber \\
\label{eq:MassOct_phase_identical-NS-form} 
\end{eqnarray}
The binary mass octupole tidal deformability $\tilde{\Lambda}^{(3)}$ is obtained as follows.
By rewriting the leading terms of Eq.~(\ref{eq:MassOct_phase}) as 
\begin{eqnarray}
&& \mathrm{(the~leading~term~of~Eq.}~(\ref{eq:MassOct_phase})) \big/ \left( \frac{3}{128\eta} x^{5/2} \right) \nonumber \\
&& \times \Lambda_A^{(3)} X_A^4 \left[ -\frac{4000}{9} (1- X_A) \right] + (A \leftrightarrow B) \nonumber \\
&& = \left( -\frac{250}{9} \right) \left( -\frac{250}{9} \right)^{-1} \left( -\frac{4000}{9} \right) \left[ (1- X_A) X_A^4 \Lambda_A^{(3)} \right. \nonumber \\
&& \left. + (A \leftrightarrow B) \right] \nonumber \\
&& = \left( -\frac{250}{9} \right) 16 \left[ (1 - X_A) X_A^4 \Lambda_A^{(3)} + (A \leftrightarrow B) \right], 
\label{eq:MassOct_phase_identical-NS-form_derive} 
\end{eqnarray}
and comparing it with Eq.~(\ref{eq:MassOct_phase_identical-NS-form}), it is obtained as 
\begin{eqnarray}
 \tilde{\Lambda}^{(3)} = 16 \left[ 
 \left( 1 - X_A \right) X_A^4 \Lambda_A^{(3)} + (A \leftrightarrow B)
\right],
\end{eqnarray}
which is the same definition as in Ref. \cite{Landry:2018bil}.
$\tilde{\Lambda}^{(3)}$ is a mass-weighted linear combination of the mass octupole component parameters $\Lambda_{A,B}^{(3)}$.

\begin{figure}[htbp]
  \begin{center}
    \includegraphics[keepaspectratio=true,height=65mm]{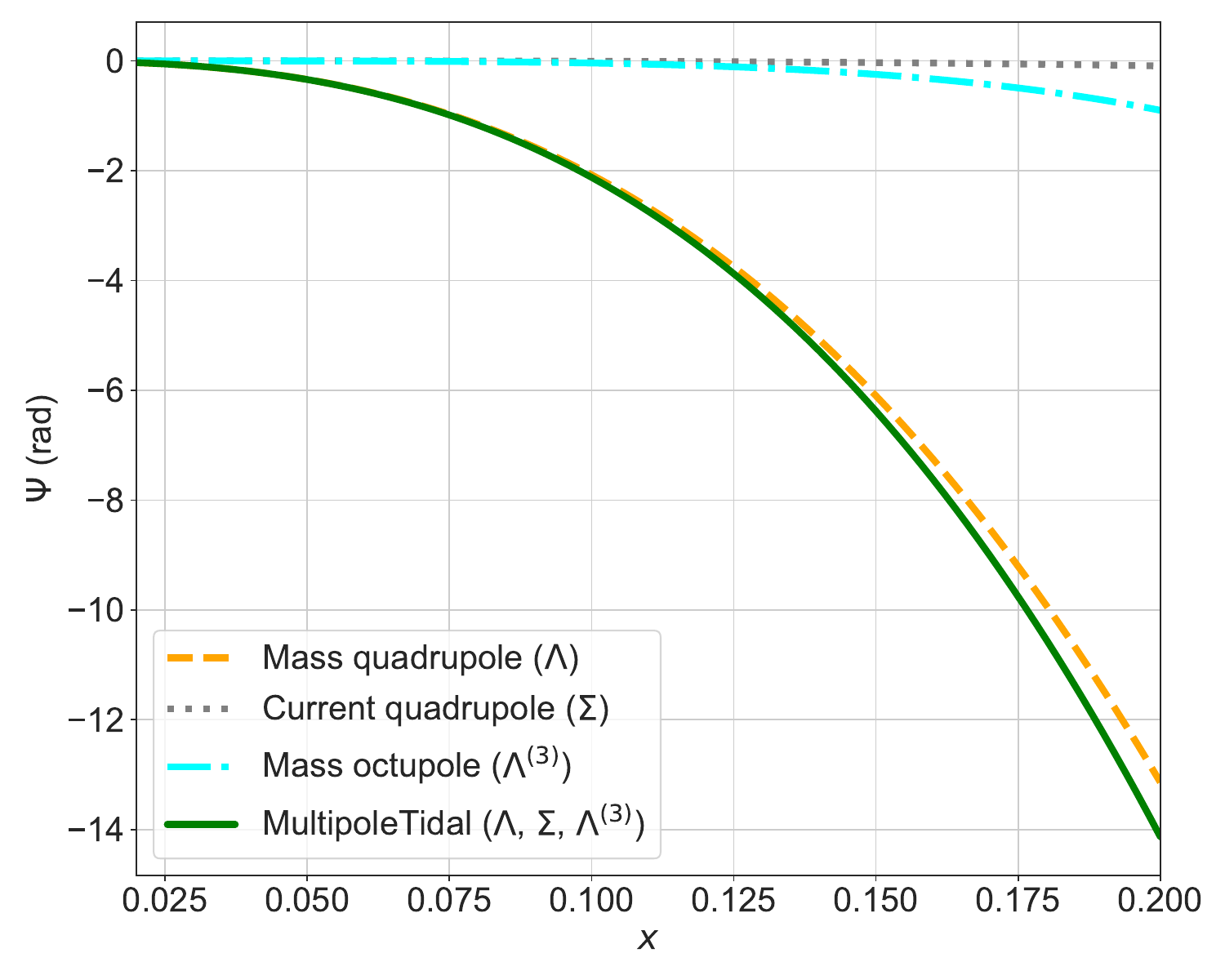}\\
  \caption{
 The phase evolution of the PN tidal phases 
 as a function of $x$.
We show the mass quadrupole $\Lambda$ (PNTidal, orange, dashed), 
the current quadrupole $\Sigma$ (gray, dotted), 
the mass octupole $\Lambda^{(3)}$ (cyan, dot-dashed), 
and the tidal multipole contributions with $\Lambda$, $\Sigma$, and $\Lambda^{(3)}$ (MultipoleTidal, green, solid)
for the equal mass BNS with $m_{A,B}=1.35~\solM$, $\Lambda_{A,B}=300$, $\Sigma_{A,B}=3.1$, and $\Lambda_{A,B}^{(3)}=483$.
}
\label{fig:Phase_AllTidal_EqualMass}
\end{center}
\end{figure}

\begin{figure}[htbp]
  \begin{center}
    \includegraphics[keepaspectratio=true,height=65mm]{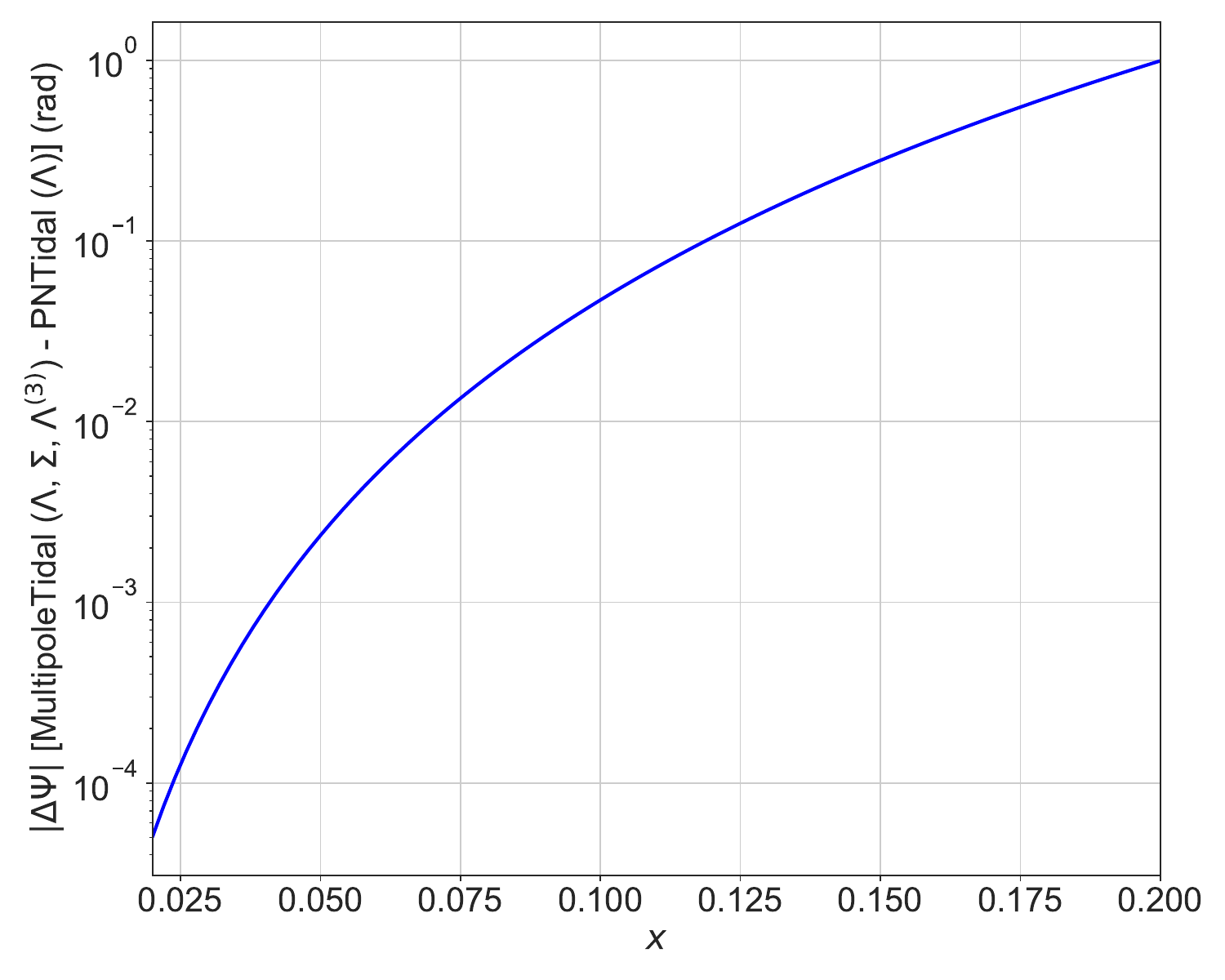}\\
  \caption{
 The absolute magnitude of the phase difference between the multipole moments with $\Lambda$, $\Sigma$, and $\Lambda^{(3)}$ (MultipoleTidal) and the mass quadrupole moment $\Lambda$ contributions (PNTidal) as a function of $x$ for the same equal mass BNS as used in Fig. \ref{fig:Phase_AllTidal_EqualMass}.
About $0.1~\mathrm{rad}$ at $1000~\mathrm{Hz}$ ($x \sim 0.125$).
}
\label{fig:PhaseDiff_AllTidal_EqualMass}
\end{center}
\end{figure}

\begin{figure}[htbp]
  \begin{center}
    \includegraphics[keepaspectratio=true,height=65mm]{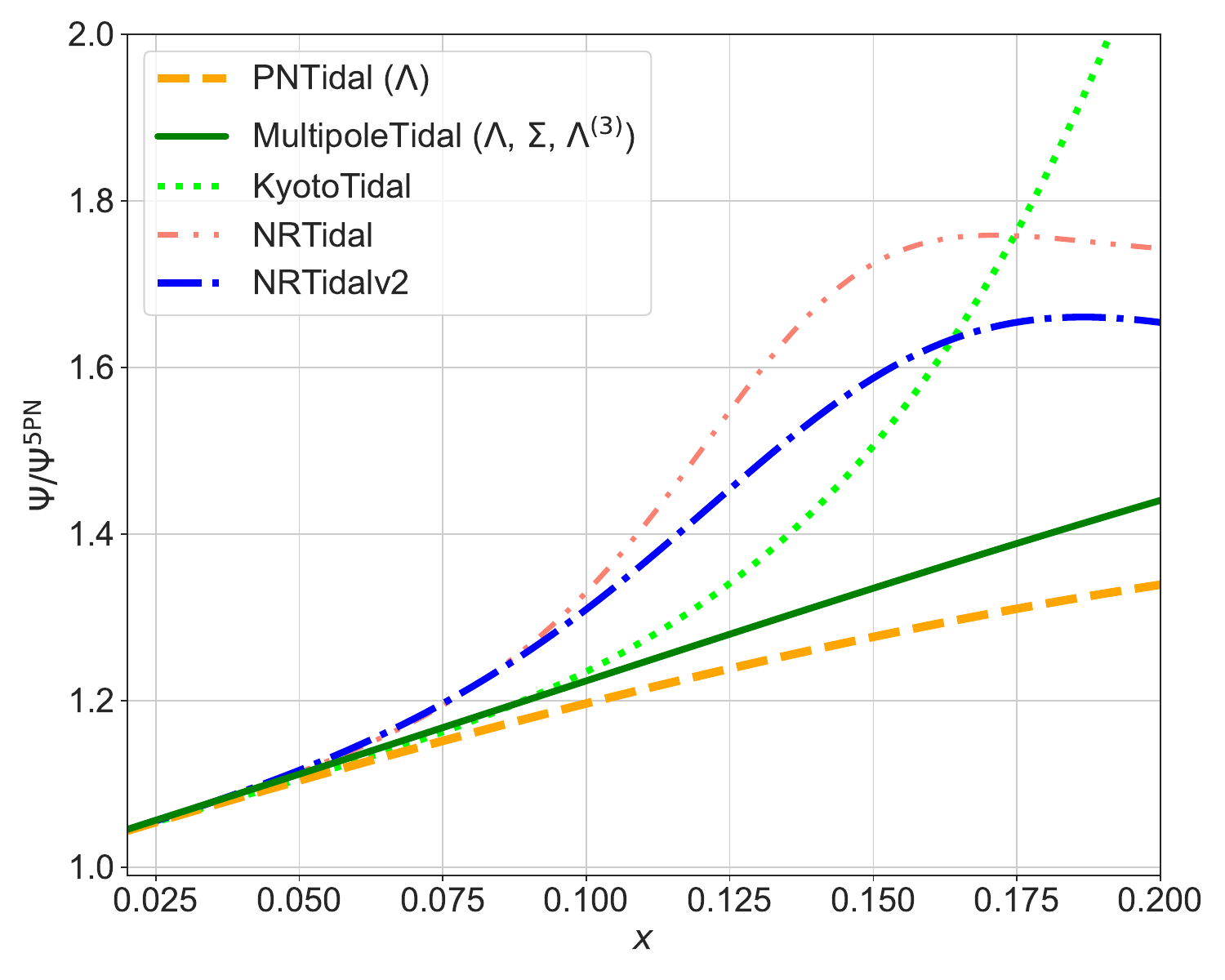}\\
  \caption{
 The phase evolution 
 for different tidal phase models normalized by the leading-order PNTidal ($\Lambda$) at the relative 5PN as a function of $x$.
We show the PNTidal ($\Lambda$, orange, dashed), the MultipoleTidal ($\Lambda$, $\Sigma$, $\Lambda^{(3)}$, green, solid), KyotoTidal (lime, dotted), NRTidal (salmon, dot-dot-dashed), and NRTidalv2 (blue, dot-dashed)
 for the same equal-mass BNS as used in Fig. \ref{fig:Phase_AllTidal_EqualMass}.
}
\label{fig:Phase_NRTidalv2_EqualMass}
\end{center}
\end{figure}

\section{Comparing MultipoleTidal with PNTidal and NRTidalv2}
\label{sec:Comparison}
In this section, 
we compare the MultipoleTidal with the PNTidal 
and the NRTidalv2 as a representative of the NR calibrated waveform model\footnote{All results shown in this paper have been obtained with the previous expression of the tidal phase, which is described in the previous version of this paper. The modification to the tidal phase Eqs.~(\ref{eq:MassQuad_phase}) and (\ref{eq:MassQuad_phase_identical-NS-form}) is numerically small. The phase difference between the previous and modified versions of the tidal phase Eq.~(\ref{eq:MassQuad_phase}) is less than $\mathcal{O}(10^{-3})$ (rad) up to 1000~Hz for a binary neutron star with $m_A=1.68~M_\odot$, $m_B=1.13~M_\odot$, $\Lambda_A=102$, and $\Lambda_B=840$.
The difference is less than $\mathcal{O}(10^{-2})$ (rad) up to 200 Hz for a binary exotic compact object with $m_A=13~M_\odot$, $m_B=6.6~M_\odot$, $\Lambda_A=-700$, and $\Lambda_B=-1400$.
Therefore, their impact on results in this paper is negligible.}.
The NRTidalv2 model is the upgrade of the NRTidal model.
We show the phase evolution for the MultipoleTidal and other models, 
the phase difference between the MultipoleTidal and the PNTidal,
the match between the MultipoleTidal and the NRTidalv2,
the one between the PNTidal and the NRTidalv2,
and the one between the MultipoleTidal and the PNTidal
for the Einstein Telescope (ET) sensitivity.
Finally, we investigate impact of the current quadrupole and the mass octupole on the parameter estimation of BNS coalescence events GW170817 and GW190425.

\subsection{Comparison of phase evolution with NR calibrated model}
\label{subsec:ComparisonPhase}
In this subsection, we compare the phase evolution of the tidal multipole moments.
Figure \ref{fig:Phase_AllTidal_EqualMass} shows the phase evolution for 
the mass quadrupole $\Lambda$ [PNTidal, Eq. (\ref{eq:MassQuad_phase})],
the current quadrupole $\Sigma$ [Eq. (\ref{eq:CurrentQuad_phase})],
the mass octupole $\Lambda^{(3)}$ [Eq. (\ref{eq:MassOct_phase})],
and the combined tidal multipole contributions (MultipoleTidal)
as a function of $x$.
Here, we consider the equal mass BNS with $m_{A,B}=1.35~\solM$, $\Lambda_{A,B}=300$, $\Sigma_{A,B}=3.1$, and $\Lambda_{A,B}^{(3)}=483$.
We use quasiuniversal fitting relations to obtain $\Sigma$ and $\Lambda^{(3)}$ from $\Lambda$ \cite{Yagi:2013sva}. 
The phase evolution for the PNTidal is very close to the one for the MultipoleTidal.
It means that the mass quadrupole contribution dominates over the multipole tidal phases.
The MultipoleTidal gives a larger phase shift than the PNTidal as x increases.

Figure \ref{fig:PhaseDiff_AllTidal_EqualMass} shows the absolute magnitude of the phase difference between the MultipoleTidal ($\Lambda$, $\Sigma$, $\Lambda^{(3)}$) and the PNTidal ($\Lambda$) as a function of $x$.
Here, we consider the same equal mass BNS as used in Fig. \ref{fig:Phase_AllTidal_EqualMass}.
The phase difference between the MultipoleTidal and the PNTidal models is about $0.1~\mathrm{rad}$ at $1000~\mathrm{Hz}$, which corresponds to $x\sim0.125$.

Figure \ref{fig:Phase_NRTidalv2_EqualMass} shows the phase evolution 
for different tidal phase models normalized by the leading-order PNTidal at the relative 5PN as a function of $x$.
We compare the PNTidal ($\Lambda$), the MultipoleTidal ($\Lambda$, $\Sigma$, $\Lambda^{(3)}$), and NRTidalv2 as a representative of the NR calibrated model
for the same equal mass BNS as used in Fig. \ref{fig:Phase_AllTidal_EqualMass}. 
The MultipoleTidal is closer to the NRTidalv2 than the PNTidal.
The NRTidalv2 gives a larger phase shift than the MultipoleTidal as $x$ increases.
The phase difference between the MultipoleTidal and the NRTidalv2 becomes larger as $x$ becomes larger up to $x\sim0.16$.
For comparison, the KyotoTidal and NRTidal are also plotted.
Here, the KyotoTidal, the NRTidal, and NRTidalv2 models are NR calibrated tidal models (the family of waveforms with tidal interactions is summarized in, e.g., Refs. \cite{Narikawa:2019xng,Dietrich:2020eud,Isoyama:2020lls}).

\subsection{Match computations with NR calibrated model}
\label{subsec:Match}
To investigate how the MultipoleTidal model differs from the PNTidal model, we compare the MultipoleTidal model, the PNTidal model, and the NRTidalv2 model by computing the match between them.
The match between waveform models is defined as the inner product between waveform models maximized over the phase and the time as
\begin{eqnarray}
 F = \underset{\phi_c,t_c}{\mathrm{max}} \frac{(\tilde{h}_1(f; \phi_c,t_c) | \tilde{h}_2(f))}{\sqrt{(\tilde{h}_1(f) | \tilde{h}_1(f) ) (\tilde{h}_2(f) | \tilde{h}_2(f) )}},
\end{eqnarray}
where $\tilde{h}_{1,2}(f)$ represents the waveform model, tildes denote the Fourier transform, and $\phi_c$, $t_c$ are an arbitrary phase and time shift.
Here, the noise-weighted inner product between waveform models is given by
\begin{eqnarray}
 (\tilde{h}_1(f) | \tilde{h}_2(f)) = 4 \mathcal{R} \int_{f_\mathrm{low}}^{f_\mathrm{high}} \frac{\tilde{h}^*_1(f)  \tilde{h}_2(f)}{S_n(f)} df,
\end{eqnarray}
where 
$S_n(f)$ is the detector noise spectral density.
We use the ET noise curve of Refs. \cite{Punturo:2010zz, P1600143} for match computation with $f_\mathrm{low}=10~\mathrm{Hz}$, since third generation GW detectors such as ET will be sensitive in the high-frequency regime above a few hundred Hz, which is suitable for observing BNS coalescence signals \cite{Branchesi:2023mws, Puecher:2023twf}.
To restrict to the inspiral regime, we set the upper frequency cutoff $f_\mathrm{high}=1024~\mathrm{Hz}$.
%
%
We use the open-source PyCBC toolkit \cite{PyCBC,Usman:2015kfa} to compute the match  for given waveform models implemented in the LALSimulation and a set of parameters.
%
To focus on the tidal phase difference, we use the same point-particle baseline, the TF2 model up to 3.5PN order for the phase, and neglect tidal amplitude contributions for all tidal models used in this study.
Here, TF2 is the abbreviation of TaylorF2, which is the PN waveform model for a point-particle part~\cite{Dhurandhar:1992mw, Buonanno:2009zt, Blanchet:2013haa} (very recently, the formula for the phase is updated to 4.5PN order \cite{Blanchet:2023bwj}).
We newly implement the TF2\_MultipoleTidal and the TF2\_NRTidalv2 in the LALSimulation. 
We use quasiuniversal fitting relations between the tidal multipole moments to obtain $\Sigma$ and $\Lambda^{(3)}$ from $\Lambda$ \cite{Yagi:2013sva} (see Appendix \ref{sec:EOS-insensitive-relations} for details). 

We select 2000 samples for BNS coalescence signals with the uniform distributions in $m_{A} \in [1,~3]~M_\odot$, $m_B \in [1,~3]~M_\odot$, and $\Lambda_{A,B} \in [20,~3000]$, imposing the mass ratio $q=m_B/m_A \leq 1$. 
Here, we consider nonspinning binaries.
%
Figure \ref{fig:McLamt_Match_ET} shows the match between the TF2\_MultipoleTidal and the TF2\_NRTidalv2 (left panel), the one between the TF2\_PNTidal and the TF2\_NRTidalv2 (middle panel), and the one between the TF2\_MultipoleTidal and the TF2\_PNTidal (right panel) on $\mathcal{M}$-$\tilde{\Lambda}$ plane.
Here, $\mathcal{M}:=(m_1 m_2)^{3/5}/(m_1 + m_2)^{1/5}$ is the chirp mass, which gives the leading-order evolution of the binary amplitude and phase.
For high chirp mass $\mathcal{M}$ and large binary tidal deformability $\tilde{\Lambda}$, the values of the match between the TF2\_MultipoleTidal and the TF2\_PNTidal models are small, corresponding to upper right side in the right panel.
The match between the TF2\_MultipoleTidal and TF2\_NRTidalv2 (left panel) is better than the one between the TF2\_PNTidal and the TF2\_NRTidalv2 (middle panel) for high $\mathcal{M}$ and large $\tilde{\Lambda}$,
where the MultipoleTidal is effectively improved by the current quadrupole $\Sigma$ and the mass octupole $\Lambda^{(3)}$ contributions.

\subsection{Analysis of GW170817 and GW190425}
\label{subsec:GW170817}
To see the impact of the multipole tidal effects on parameter estimation,
we apply the MultipoleTidal model to BNS coalescence events GW170817 and GW190425
using the parameter estimation software, LALInference \cite{Veitch:2014wba, LAL}.
We analyze 128 seconds data around coalescence time and frequency range between 23 and 1000 Hz for GW170817 and between 19.4 and 1000 Hz for GW190425.
We compare three tidal phase models:
the MultipoleTidal model (combination of Eq. (\ref{eq:MassQuad_phase}) for the mass quadrupole $\Lambda$, Eq. (\ref{eq:CurrentQuad_phase}) for the current quadrupole $\Sigma$, and Eq. (\ref{eq:MassOct_phase}) for the mass octupole $\Lambda^{(3)}$),
the PNTidal model (Eq. (\ref{eq:MassQuad_phase}) for the mass quadrupole $\Lambda$), 
and the NRTidalv2 model for the NR calibrated effects.
Here, we do not consider the tidal amplitude contribution for all tidal models to focus on the difference in the phases.
We use the TF2 model 
as the inspiral point-particle baseline for all tidal models.
We assume that the spins of component objects are aligned with the orbital angular momentum 
and incorporate the 3.5PN order formula in couplings between the orbital angular momentum 
and the component spins~\cite{Bohe:2013cla},
3PN order formula in point-mass spin-spin, 
and self-spin interactions~\cite{Arun:2008kb, Mikoczi:2005dn}.

We are basically following our previous analyses of GW170817 and GW190425~\cite{Narikawa:2022saj}, except for the waveform models used.
We compute posterior probability distribution functions (PDFs) and Bayes factor (BF) using the nested sampling algorithm \cite{Skilling:2004,Skilling:2006,Ashton:2022grj} available in the LALInference package \cite{Veitch:2014wba} as part of the LAL \cite{LAL}.
The data are available on the Gravitational Wave Open Science Center
released by the LIGO-Virgo-KAGRA (LVK) Collaborations~\cite{LIGOScientific:2019lzm}.
The noise spectrum densities estimated with the BayesLine algorithm \cite{Cornish:2014kda, Littenberg:2015kpb, Chatziioannou:2019zvs} are also obtained from there.
We employ a uniform prior on the detector-frame component masses $m_{1,2}^\mathrm{det}$ in the range $[0.5,~5.0]M_\odot$,
the spin magnitudes $\chi_{1z,2z}$ in the range $[-0.05,~0.05]$,
the mass quadrupole tidal deformabilities $\Lambda_{1}$ in the range $[0, 5000]$ and $\Lambda_{2}$ in the range $[0, 10000]$.
To obtain $\Sigma$ and $\Lambda^{(3)}$ from $\Lambda$,
we use quasiuniversal fitting relations between the tidal multipole moments \cite{Yagi:2013sva} (see Appendix \ref{sec:EOS-insensitive-relations} for details).
%
In our analyses, we marginalized over the coalescence time $t_c$ and the phase at the coalescence time $\phi_c$ semianalytically.
For GW170817, we fix the sky location to the position of AT 2017gfo, which is an electromagnetic counterpart of GW170817~\cite{Soares-Santos:2017lru, LIGOScientific:2017ync, J-GEM:2017tyx}.

Figure \ref{fig:Lamt_GW170817_GW190425} shows the marginalized posterior PDFs of the binary tidal deformability $\tilde{\Lambda}$ for GW170817 (left) and GW190425 (right) using TF2\_MultipoleTidal, TF2\_PNTidal, and TF2\_NRTidalv2 models.
%
The corresponding 90\% credible intervals [highest-probability-density (HPD)],
, logarithmic Bayes factor of the signal hypothesis against the noise assumption for each model,
and logarithmic Bayes factors between three tidal phase models and the TF2\_MultipoleTidal model are summarized in Table \ref{table:logBF_GW170817_GW190425}.
Here, to effectively obtain a uniform prior on $\tilde{\Lambda}$, we weight the posterior of $\tilde{\Lambda}$ by dividing by the prior, similarly to the LVC's analyses \cite{TheLIGOScientific:2017qsa, Abbott:2018wiz}.
The difference between the inferred $\tilde{\Lambda}$ by TF2\_MultipoleTidal and TF2\_PNTidal is very small compared with 90\% statistical error.
The impact of the additional tidal effects in the MultipoleTidal models is not significant on the estimates of $\tilde{\Lambda}$, which is consistent with the results of Ref. \cite{Pradhan:2022rxs}.
A closer look reveals that the TF2\_NRTidalv2 gives a smallest median value and narrowest 90\% credible interval for the inferred $\tilde{\Lambda}$.
The TF2\_MultipoleTidal gives smaller median value and narrower 90\% credible interval for the inferred $\tilde{\Lambda}$ than the TF2\_PNTidal, and is closer distribution to the TF2\_NRTidalv2 than the TF2\_PNTidal.
These results are expected by the tidal phase shift shown in Fig. \ref{fig:Phase_NRTidalv2_EqualMass}.
The logarithmic Bayes factor of the TF2\_PNTidal and the TF2\_NRTidalv2 relative to the TF2\_MultipoleTidal,
$\log\mathrm{BF}^\mathrm{X}_\mathrm{TF2\_MultipoleTidal}$, 
where $X=\mathrm{TF2\_PNTidal}$ or $\mathrm{TF2\_NRTidalv2}$, 
indicate no preference among the three tidal phase models by relying on BNS coalescences GW170817 and GW190425 as shown in Table \ref{table:logBF_GW170817_GW190425}.

\begin{widetext}

\begin{figure*}[htbp]
  \begin{center}
\begin{tabular}{lcr}
 \begin{minipage}[h]{0.3\linewidth}
 \begin{center}
    \includegraphics[keepaspectratio=true,height=41mm]{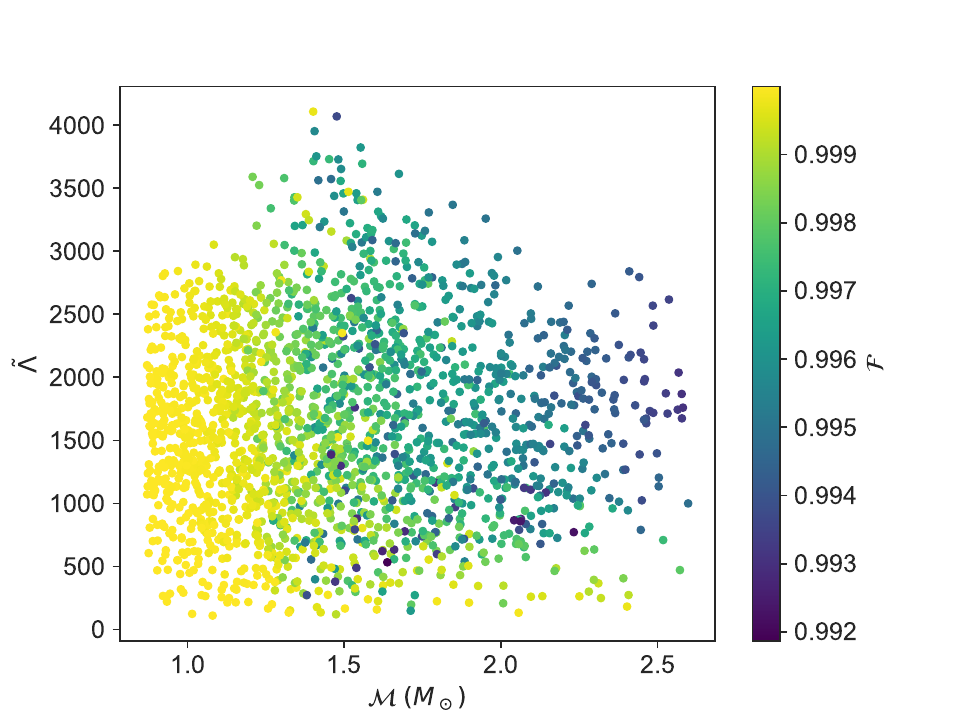}\\
 \end{center}
 \end{minipage}
 \begin{minipage}[h]{0.3\linewidth}
  \begin{center}
    \includegraphics[keepaspectratio=true,height=41mm]{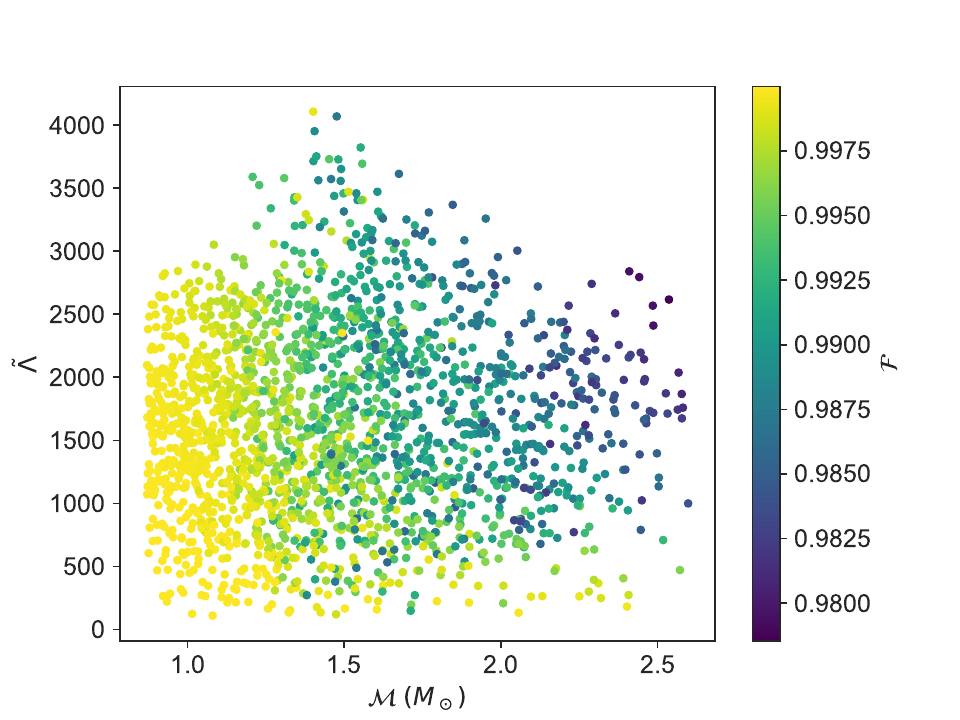}\\
     \end{center}
 \end{minipage}
 \begin{minipage}[h]{0.3\linewidth}
  \begin{center}
    \includegraphics[keepaspectratio=true,height=41mm]{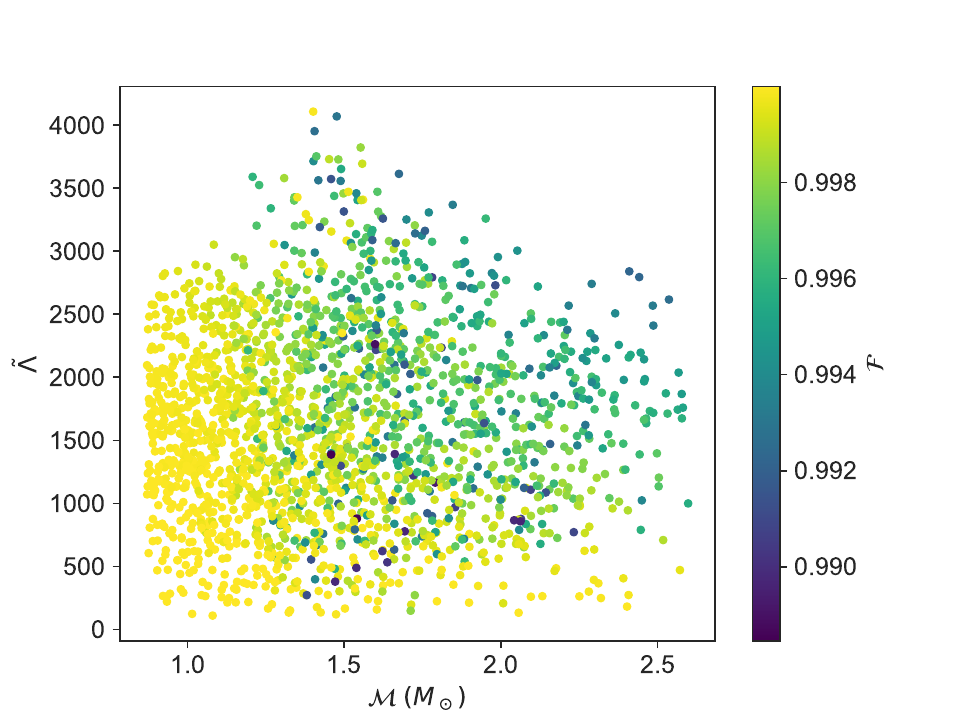}
     \end{center}
 \end{minipage}
\end{tabular}
  \caption{
 The match between the TF2\_MultipoleTidal and TF2\_NRTidalv2 (left), the one between the TF2\_PNTidal and TF2\_NRTidalv2 (middle), and the one between the TF2\_MultipoleTidal and TF2\_PNTidal (right) on $\mathcal{M}$-$\tilde{\Lambda}$ plane.
 }%
\label{fig:McLamt_Match_ET}
\end{center}
\end{figure*}

\begin{figure*}[htbp]
  \begin{center}
\begin{tabular}{cc}
 \begin{minipage}[b]{0.45\linewidth}
 \begin{center}
    \includegraphics[keepaspectratio=true,height=80mm]{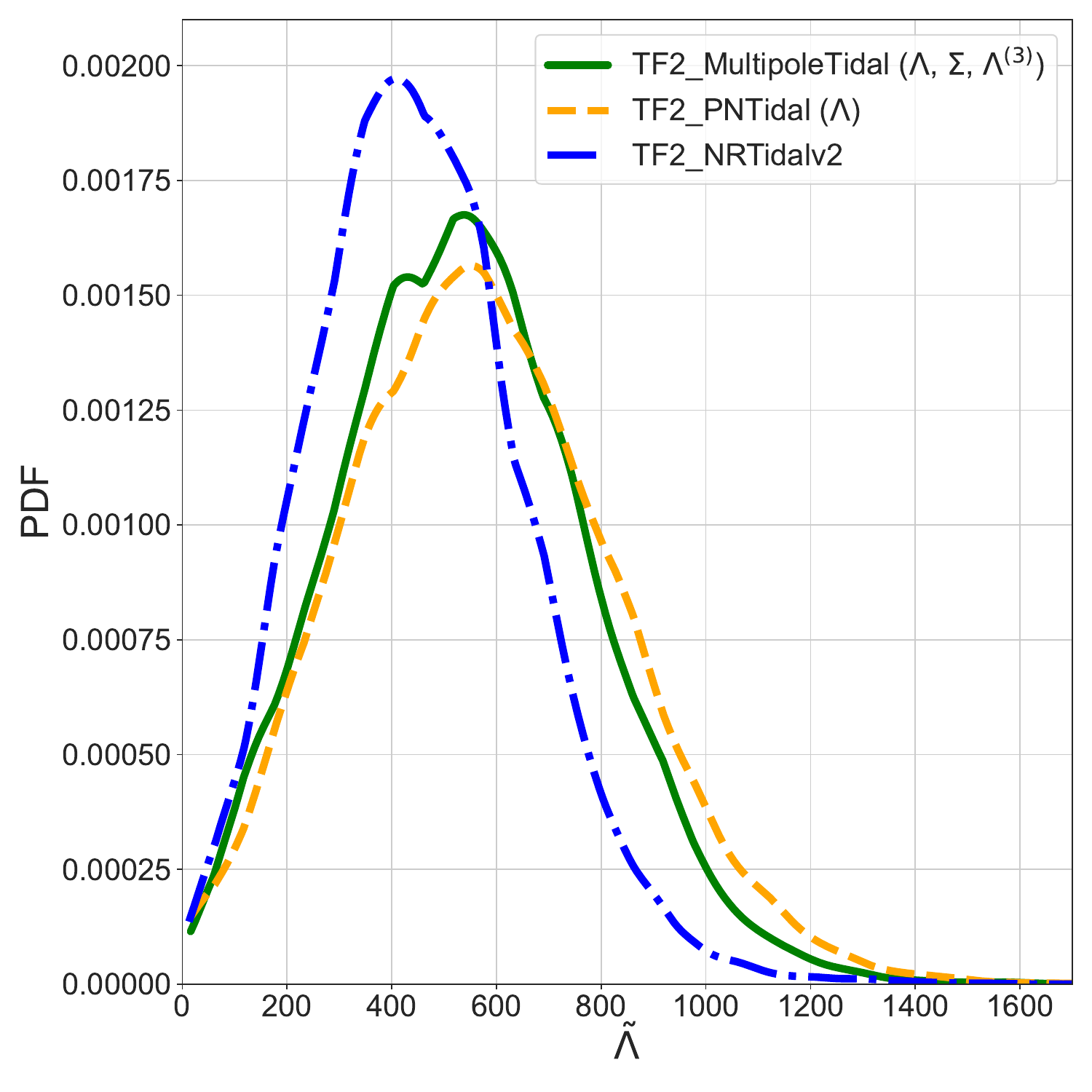}\\
 \end{center}
 \end{minipage}
 \begin{minipage}[b]{0.45\linewidth}
  \begin{center}
    \includegraphics[keepaspectratio=true,height=80mm]{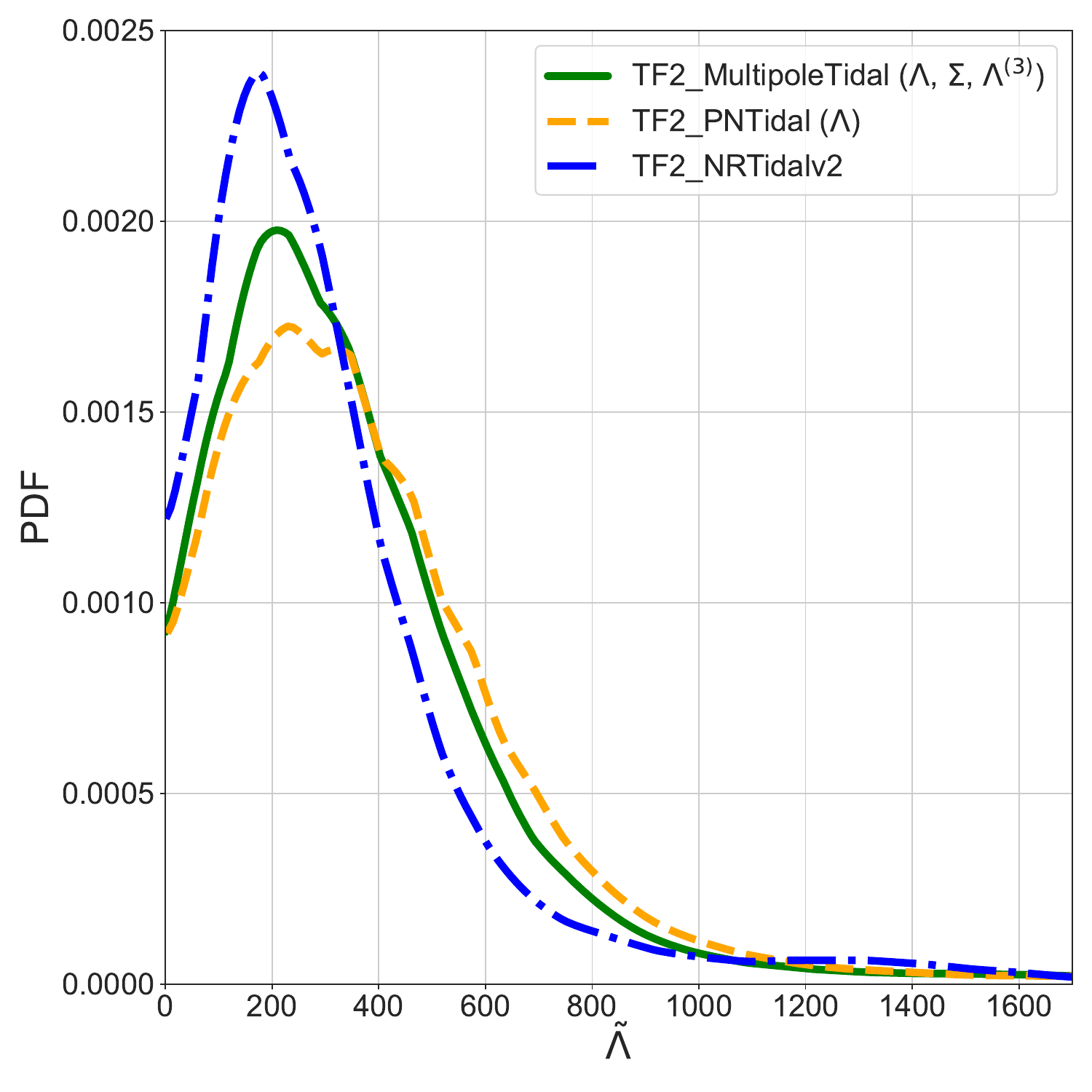}\\
 \end{center}
 \end{minipage}
\end{tabular}
  \caption{
Marginalized posterior PDFs of the binary tidal deformability $\tilde{\Lambda}$ 
for GW170817 (left) and GW190425 (right) 
estimated by using three tidal phase models, employing the same point-particle model, TF2,
for $|\chi_{1z,2z}| \leq 0.05$ and the upper frequency cutoff $f_\mathrm{high}=1000~\mathrm{Hz}$.
The curves correspond to 
the TF2\_MultipoleTidal (green, solid), TF2\_PNTidal (orange, dashed), and TF2\_NRTidalv2 (blue, dot-dashed) models.
The corresponding 90\% credible intervals (HPD) and logarithmic Bayes factors between three tidal phase models and the TF2\_MultipoleTidal model are presented in Table \ref{table:logBF_GW170817_GW190425}.
}
\label{fig:Lamt_GW170817_GW190425}
\end{center}
\end{figure*}

\begin{table}[htbp]
\begin{center}
\caption{
The 90\% credible intervals (HPD) for $\tilde{\Lambda}$, the logarithmic Bayes factor of the signal hypothesis against the noise assumption for each tidal model, 
and the logarithmic Bayes factor 
of the TF2\_PNTidal and the TF2\_NRTidalv2 relative to the TF2\_MultipoleTidal,
$\log\mathrm{BF}^\mathrm{X}_\mathrm{TF2\_MultipoleTidal}$, 
for GW170817 and GW190425. 
The values indicate no preference model on GW170817 and GW190425.
}
\vspace{5pt}
\begin{tabular}{lccccccc}
\hline \hline
 & ~~GW170817 & & &  ~~GW190425 \\ \hline
Waveform (X) & $\tilde{\Lambda}$ & $\log \mathrm{BF}^\mathrm{S}_\mathrm{N}$ & $\log \mathrm{BF}^\mathrm{X}_\mathrm{TF2\_MultipoleTidal}$ & $\tilde{\Lambda}$ & $\log \mathrm{BF}^\mathrm{S}_\mathrm{N}$ & $\log \mathrm{BF}^\mathrm{X}_\mathrm{TF2\_MultipoleTidal}$ \\ \hline
TF2\_MultipoleTidal & $539_{-405}^{+375}$ & $503.0$ & 0 & $209_{-209}^{+469}$ & 46.6 & 0 \\
TF2\_PNTidal & $551_{-406}^{+429}$ & $503.0$ & 0.0 & $229_{-229}^{+505}$ & 46.9 & 0.3  \\
TF2\_NRTidalv2 & $410_{-292}^{+360}$ & $502.7$ & -0.3 & $174_{-174}^{+449}$ & 46.3 & -0.3 \\
\hline \hline
\end{tabular}
\label{table:logBF_GW170817_GW190425}
\end{center}
\end{table}

\end{widetext}

\section{Conclusion}
\label{sec:summary}
PN GW phases of compact binary coalescences for the tidal multipole moments, which consist of the mass quadrupole $\Lambda$, the current quadrupole $\Sigma$, and the mass octupole moments $\Lambda^{(3)}$, have been completed up to 5+2.5PN order in Ref. \cite{Henry:2020ski}.
We rewrite the original form to the multipole component form as a function of the multipole component tidal deformabilities, which is convenient for data analysis.
To see the impact of the current quadrupole and mass octupole moments, we compare the MultipoleTidal with the PNTidal as well as NRTidalv2 by computing the phase evolution, the phase difference, the matches between waveform models, and applying parameter estimation for BNS coalescence events GW170817 and GW190425.

First, comparing the phase evolution for different tidal waveform models shows that the MultipoleTidal gives a larger phase shift than the PNTidal, and is closer to the NRTidalv2.
The phase difference between the MultipoleTidal and the PNTidal is about $0.1~\mathrm{rad}$ at $1000~\mathrm{Hz}$.

Second, comparing matches between different waveform models provides
that
the MultipoleTidal is closer to the NRTidalv2 than the PNTidal, in particular, for high masses and large tidal deformabilities.

Finally, we compare parameter estimation results from different waveform models for BNS coalescence events GW170817 and GW190425.
We find that the difference between the inferred binary tidal deformability $\tilde{\Lambda}$ by TF2\_MultipoleTidal and TF2\_PNTidal is very small.
It means that the additional current quadrupole and mass octupole moments give no significant systematic difference in the inferred $\tilde{\Lambda}$ compared to the mass quadrupole moment.
These results are consistent with the phase evolution and the results of Ref. \cite{Pradhan:2022rxs}.
%
The estimated logarithmic Bayes factors between tidal waveform models show no preference among the three tidal phase models for GW170817 and GW190425.

During the fourth observing run 4 (O4), 
it is expected that tens of BNS coalescence signals will be detected, which provide rich information on the sources~\cite{KAGRA:2013rdx}.
As the number of detected BNS coalescence events increases, 
the systematic differences among different tidal waveform models will be noticeable 
and constraints on EOS models for NSs will be improved
as shown in Refs.~\cite{DelPozzo:2013ala, Agathos:2015uaa, Lackey:2014fwa, Wysocki:2020myz, Dudi:2018jzn, Samajdar:2018dcx, Messina:2019uby, Samajdar:2019ulq, Agathos:2019sah, Gamba:2020wgg, Landry:2020vaw, Chen:2020fzm, Chatziioannou:2021tdi, Kunert:2021hgm}.
Therefore, coming BNS coalescence data would reveal the need for expansions from PN tidal models.
The multipole component form and the identical-NS form for the tidal multipole moments derived in this paper can be used to extend NR calibrated models including multipole tidal interactions.
Through tidal multipole moments, additional information on EOS for NSs will be obtained.

\section*{Acknowledgment}
We would like to thank Nami Uchikata, Kyohei Kawaguchi, and Hideyuki Tagoshi for fruitful discussions and useful comments on the study.
This work is supported by JSPS KAKENHI Grants No. JP21K03548.
The analyses in this paper were run on the VELA cluster in the ICRR. 
This research has made use of data or software obtained from the Gravitational Wave Open Science Center (gw-openscience.org), a service of LIGO Laboratory, the LIGO Scientific Collaboration, the Virgo Collaboration, and KAGRA. LIGO Laboratory and Advanced LIGO are funded by the United States National Science Foundation (NSF) as well as the Science and Technology Facilities Council (STFC) of the United Kingdom, the Max-Planck-Society (MPS), and the State of Niedersachsen/Germany for support of the construction of Advanced LIGO and construction and operation of the GEO\,600 detector. Additional support for Advanced LIGO was provided by the Australian Research Council. Virgo is funded, through the European Gravitational Observatory (EGO), by the French Centre National de Recherche Scientifique (CNRS), the Italian Istituto Nazionale di Fisica Nucleare (INFN) and the Dutch Nikhef, with contributions by institutions from Belgium, Germany, Greece, Hungary, Ireland, Japan, Monaco, Poland, Portugal, Spain. The construction and operation of KAGRA are funded by Ministry of Education, Culture, Sports, Science and Technology (MEXT), and Japan Society for the Promotion of Science (JSPS), National Research Foundation (NRF) and Ministry of Science and ICT (MSIT) in Korea, Academia Sinica (AS) and the Ministry of Science and Technology (MoST) in Taiwan.

\appendix
\section{EOS-insensitive relations}
\label{sec:EOS-insensitive-relations}
In this appendix, we summarize the EOS-insensitive relations for the tidal multipole moments,
which are used to obtain $\Sigma$, and $\Lambda^{(3)}$ from $\Lambda$.
A fitting function (including only the realistic EoSs) of the form is constructed as Eq. (60) of Ref. \cite{Yagi:2013sva}
\begin{eqnarray}
 \ln y_i = a_i + b_i \ln x_i + c_i (\ln x_i)^2 +d_i (\ln x_i)^3 + e_i (\ln x_i)^4, \nonumber \\
 \label{eq:EOS-fit-form}
\end{eqnarray}
where the fitted coefficients are in Table I of Ref. \cite{Yagi:2013sva} as shown in Table \ref{table:EOS-fit-coef}.
By using the EOS-insensitive relations of the tidal multipole moments,
$\Sigma=3.11$ and $\Lambda^{(3)}=483.3$ are obtained for $\Lambda=300$.

\section{Waveform systematics for the different tidal effects in the source properties}
\label{sec:WFsys}
In this appendix, we present estimates of source parameters for completeness obtained by using three different tidal models: the MultipoleTidal, PNTidal, and NRTidalv2 models, 
employing the same point-particle baseline model TF2 for $|\chi_{1z,2z}| \leq 0.05$ and the upper frequency cutoff $f_\mathrm{high}=1000~\mathrm{Hz}$.
We demonstrate that the inferred marginalized masses and spins are not sensitive to the different tidal models, unlike $\tilde{\Lambda}$. 

Figures~\ref{fig:McqChieffLamt_GW170817} and \ref{fig:McqChieffLamt_GW190425} 
show two-dimensional posterior PDFs of ($\mathcal{M}$, $q$, $\chi_\mathrm{eff}$, $\tilde{\Lambda}$) for GW170817 and GW190425, respectively.
%
%
The estimates of source parameters presented
show the absence of significant systematic difference 
associated with a difference among tidal part models: the MultipoleTidal, PNTidal, and NRTidalv2 models.

\begin{widetext}

\begin{table}[htbp]
\begin{center}
\caption{
The fitted coefficients for the fitting formula for the tidal multipole moments given in Eq. (\ref{eq:EOS-fit-form}).
The last line is updated in Errata of Ref. \cite{Yagi:2013sva}.
}
\vspace{5pt}
\begin{tabular}{lccccccc}
\hline \hline
$y_i$~~~ & ~~~~$x_i$ &  ~~~$a_i$ & ~~~$b_i$ & ~~~$c_i$ & ~~~$d_i$ & ~~~$e_i$\\ \hline
$\Lambda^{(3)}$ & ~~~$\Lambda$ & ~~~-1.15 & ~~~1.18 & ~~~2.51$\times10^{-2}$ & ~~~$-1.31\times10^{-3}$ & ~~~$2.52\times10^{-5}$ \\
$|\Sigma|$ & ~~~$\Lambda$ & ~~~-2.01 & ~~~0.462 & ~~~$1.68\times10^{-2}$ & ~~~$-1.58\times10^{-4}$ & ~~~$-6.03\times10^{-6}$ \\
\hline \hline
\end{tabular}
\label{table:EOS-fit-coef}
\end{center}
\end{table}

\begin{figure}[htbp]
  \begin{center}
 \begin{center}
    \includegraphics[keepaspectratio=true,height=160mm]{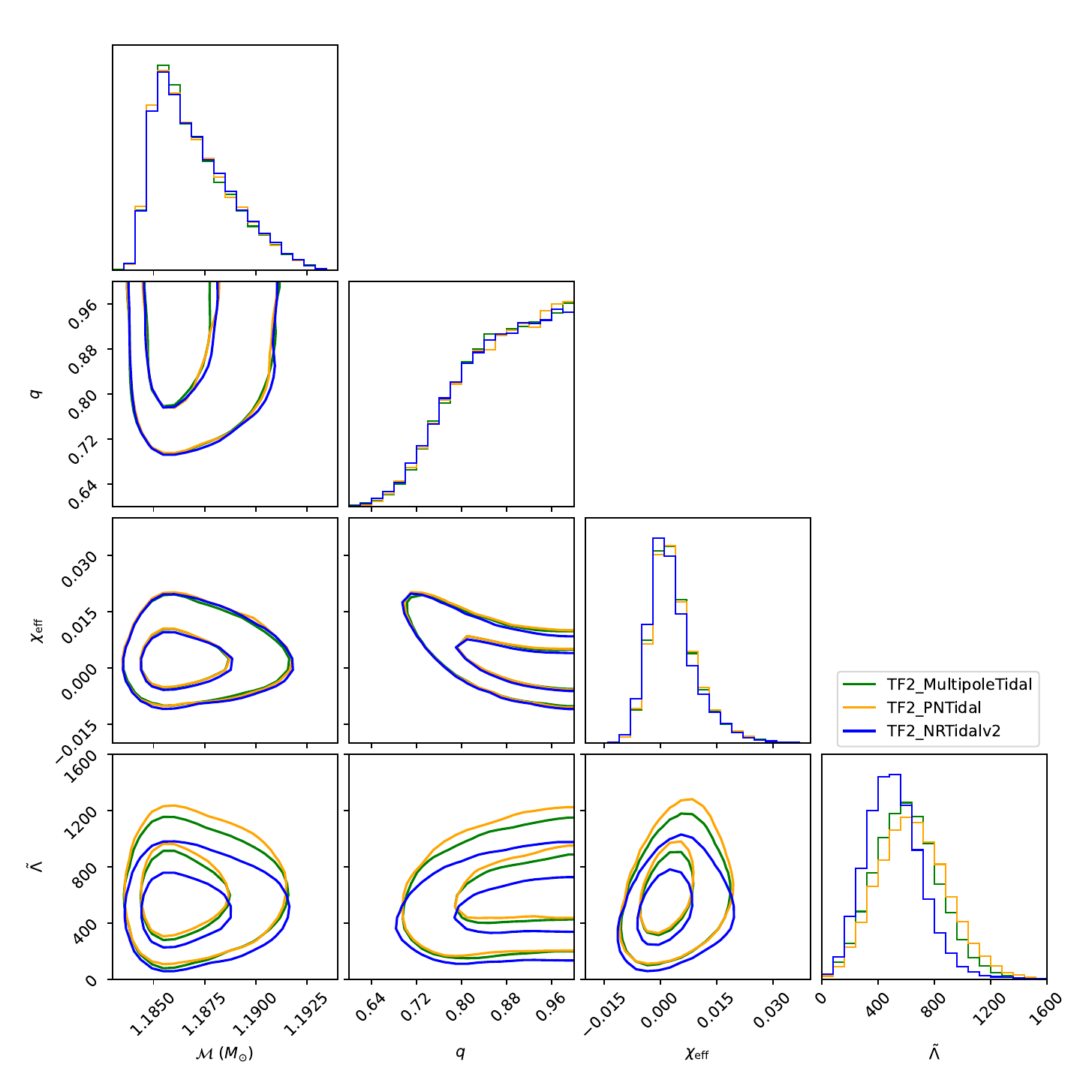}\\
 \end{center}
  \caption{
Comparison of two-dimensional posterior PDFs of ($\mathcal{M}$, $q$, $\chi_\mathrm{eff}$, $\tilde{\Lambda}$) for GW170817.
Contours show 50\% and 90\% credible regions for the TF2\_MultipoleTidal (green), TF2\_PNTidal (orange), and TF2\_NRTidalv2 (blue) models for $|\chi_{1z,2z}| \leq 0.05$ and the upper frequency cutoff $f_\mathrm{high}=1000~\mathrm{Hz}$.
The systematic difference for masses and spins associated with a difference among waveform models for the tidal part are very small, unlike $\tilde{\Lambda}$.
}%
\label{fig:McqChieffLamt_GW170817}
\end{center}
\end{figure}

\begin{figure}[htbp]
  \begin{center}
 \begin{center}
    \includegraphics[keepaspectratio=true,height=160mm]{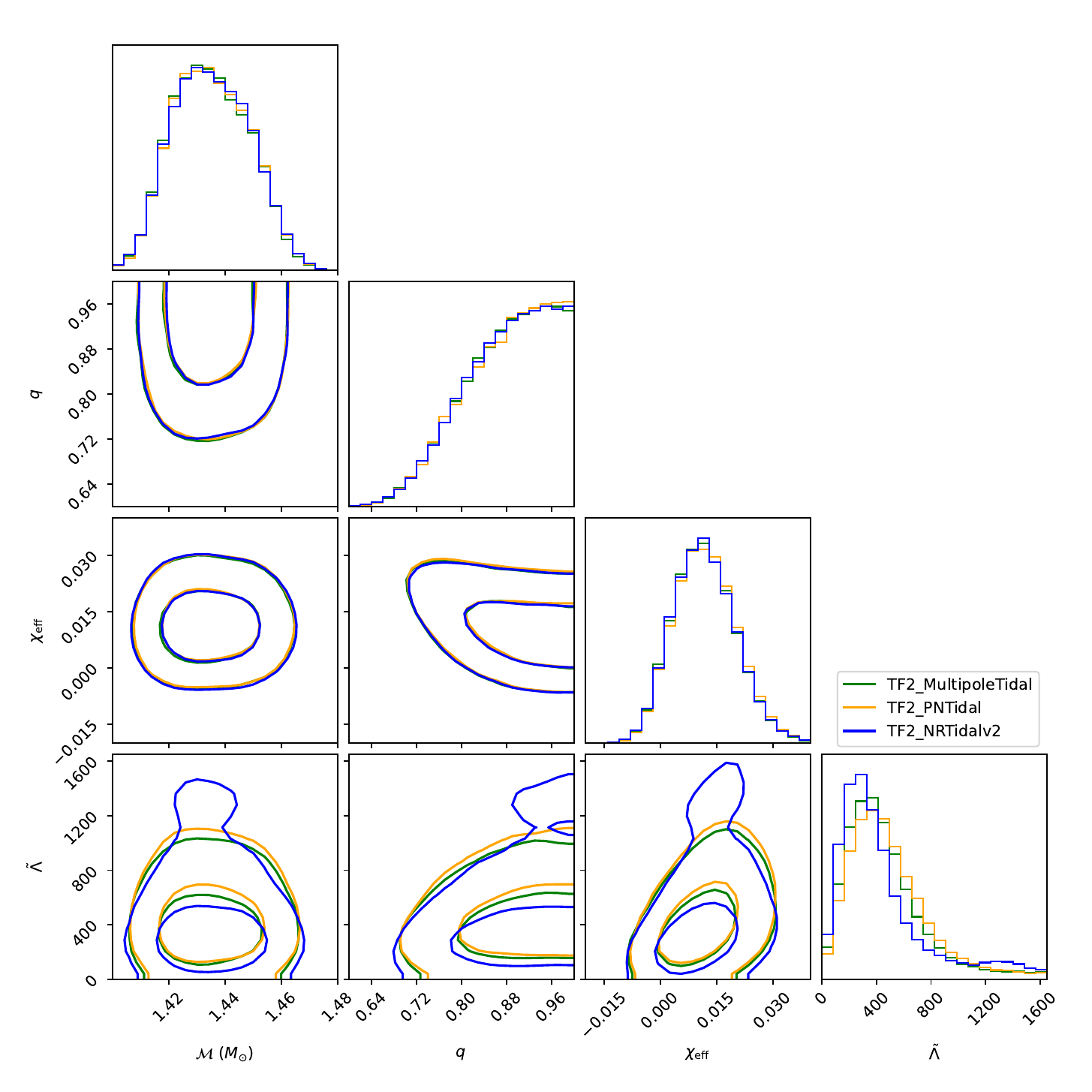}\\
 \end{center}
  \caption{
Same as Fig.~\ref{fig:McqChieffLamt_GW170817} but for GW190425.
The systematic difference for masses and spins associated with a difference among tidal waveform models are very small, unlike $\tilde{\Lambda}$.
}%
\label{fig:McqChieffLamt_GW190425}
\end{center}
\end{figure}
\end{widetext}

  


\begin{thebibliography}{99}  
\bibitem{Damour:2012yf} 
  T.~Damour, A.~Nagar and L.~Villain,
  ``Measurability of the tidal polarizability of neutron stars in late-inspiral gravitational-wave signals'',
  Phys.\ Rev.\ D {\bf 85}, 123007 (2012)
  [arXiv:1203.4352 [gr-qc]].        

\bibitem{Bini:2012gu}
D.~Bini, T.~Damour and G.~Faye,
``Effective action approach to higher-order relativistic tidal interactions in binary systems and their effective one body description,''
Phys. Rev. D \textbf{85}, 124034 (2012)
[arXiv:1202.3565 [gr-qc]].

\bibitem{Agathos:2015uaa}
M.~Agathos, J.~Meidam, W.~Del Pozzo, T.~G.~F.~Li, M.~Tompitak, J.~Veitch, S.~Vitale and C.~Van Den Broeck,
``Constraining the neutron star equation of state with gravitational wave signals from coalescing binary neutron stars,''
Phys. Rev. D \textbf{92}, 023012 (2015)
[arXiv:1503.05405 [gr-qc]].

\bibitem{Henry:2020ski}
Q.~Henry, G.~Faye and L.~Blanchet,
``Tidal effects in the gravitational-wave phase evolution of compact binary systems to next-to-next-to-leading post-Newtonian order,''
Phys. Rev. D \textbf{102}, 044033 (2020)
[arXiv:2005.13367 [gr-qc]].

\bibitem{TheLIGOScientific:2017qsa} 
  B.~P.~Abbott {\it et al.} [LIGO Scientific and Virgo Collaborations],
  ``GW170817: Observation of Gravitational Waves from a Binary Neutron Star Inspiral'',
  Phys.\ Rev.\ Lett.\  {\bf 119}, 161101 (2017)
  [arXiv:1710.05832 [gr-qc]].

\bibitem{Abbott:2018wiz} 
  B.~P.~Abbott {\it et al.} [LIGO Scientific and Virgo Collaborations],
  ``Properties of the binary neutron star merger GW170817,''
  Phys.\ Rev.\ X {\bf 9}, 011001 (2019)
  [arXiv:1805.11579 [gr-qc]].  

\bibitem{De:2018uhw} 
  S.~De, D.~Finstad, J.~M.~Lattimer, D.~A.~Brown, E.~Berger and C.~M.~Biwer,
  ``Tidal Deformabilities and Radii of Neutron Stars from the Observation of GW170817,''
  Phys.\ Rev.\ Lett.\  {\bf 121}, 091102 (2018)
  Erratum: [Phys.\ Rev.\ Lett.\  {\bf 121}, 259902 (2018)]
  [arXiv:1804.08583 [astro-ph.HE]].

\bibitem{LIGOScientific:2018mvr}
B.~P.~Abbott \textit{et al.} [LIGO Scientific and Virgo],
``GWTC-1: A Gravitational-Wave Transient Catalog of Compact Binary Mergers Observed by LIGO and Virgo during the First and Second Observing Runs,''
Phys. Rev. X \textbf{9}, 031040 (2019)
[arXiv:1811.12907 [astro-ph.HE]].

\bibitem{Abbott:2020uma}
  B.~P.~Abbott {\it et al.} [LIGO Scientific and Virgo Collaborations],
``GW190425: Observation of a Compact Binary Coalescence with Total Mass $\sim 3.4 M_{\odot}$,''
Astrophys. J. Lett. \textbf{892}, L3 (2020)
[arXiv:2001.01761 [astro-ph.HE]].

\bibitem{LIGOScientific:2020ibl}
R.~Abbott \textit{et al.} [LIGO Scientific and Virgo],
``GWTC-2: Compact Binary Coalescences Observed by LIGO and Virgo During the First Half of the Third Observing Run,''
Phys. Rev. X \textbf{11}, 021053 (2021)
[arXiv:2010.14527 [gr-qc]].

\bibitem{Dai:2018dca}
L.~Dai, T.~Venumadhav and B.~Zackay,
``Parameter Estimation for GW170817 using Relative Binning,''
[arXiv:1806.08793 [gr-qc]].

\bibitem{Narikawa:2018yzt}
T.~Narikawa, N.~Uchikata, K.~Kawaguchi, K.~Kiuchi, K.~Kyutoku, M.~Shibata and H.~Tagoshi,
``Discrepancy in tidal deformability of GW170817 between the Advanced LIGO twin detectors,''
Phys. Rev. Research. \textbf{1}, 033055 (2019)
[arXiv:1812.06100 [astro-ph.HE]].

\bibitem{Narikawa:2019xng}
T.~Narikawa, N.~Uchikata, K.~Kawaguchi, K.~Kiuchi, K.~Kyutoku, M.~Shibata and H.~Tagoshi,
``Reanalysis of the binary neutron star mergers GW170817 and GW190425 using numerical-relativity calibrated waveform models,''
Phys. Rev. Res. \textbf{2}, 043039 (2020)
[arXiv:1910.08971 [gr-qc]].

\bibitem{Gamba:2020wgg}
R.~Gamba, M.~Breschi, S.~Bernuzzi, M.~Agathos and A.~Nagar,
``Waveform systematics in the gravitational-wave inference of tidal parameters and equation of state from binary neutron star signals,''
Phys. Rev. D \textbf{103}, 124015 (2021)
[arXiv:2009.08467 [gr-qc]].

\bibitem{Breschi:2021wzr}
M.~Breschi, R.~Gamba and S.~Bernuzzi,
``Bayesian inference of multimessenger astrophysical data: Methods and applications to gravitational waves,''
Phys. Rev. D \textbf{104}, 042001 (2021)
[arXiv:2102.00017 [gr-qc]].

\bibitem{Narikawa:2022saj}
T.~Narikawa and N.~Uchikata,
``Follow-up analyses of the binary-neutron-star signals GW170817 and GW190425 by using post-Newtonian waveform models,''
Phys. Rev. D \textbf{106}, 103006 (2022)
[arXiv:2205.06023 [gr-qc]].

\bibitem{LIGOScientific:2018ehx}
B.~P.~Abbott \textit{et al.} [LIGO Scientific and Virgo],
``Constraining the $p$-Mode\textendash{}$g$-Mode Tidal Instability with GW170817,''
Phys. Rev. Lett. \textbf{122}, 061104 (2019)
[arXiv:1808.08676 [astro-ph.HE]].

\bibitem{Pratten:2019sed}
G.~Pratten, P.~Schmidt and T.~Hinderer,
``Gravitational-Wave Asteroseismology with Fundamental Modes from Compact Binary Inspirals,''
Nature Commun. \textbf{11}, 2553 (2020)
[arXiv:1905.00817 [gr-qc]].

\bibitem{Pan:2020tht}
Z.~Pan, Z.~Lyu, B.~Bonga, N.~Ortiz and H.~Yang,
``Probing Crust Meltdown in Inspiraling Binary Neutron Stars,''
Phys. Rev. Lett. \textbf{125}, no.20, 201102 (2020)
[arXiv:2003.03330 [astro-ph.HE]].

\bibitem{Pradhan:2022rxs}
B.~K.~Pradhan, A.~Vijaykumar and D.~Chatterjee,
``Impact of updated multipole Love numbers and f-Love universal relations in the context of binary neutron stars,''
Phys. Rev. D \textbf{107}, 023010 (2023)
[arXiv:2210.09425 [astro-ph.HE]].

\bibitem{Narikawa:2021pak}
T.~Narikawa, N.~Uchikata and T.~Tanaka,
``Gravitational-wave constraints on the GWTC-2 events by measuring the tidal deformability and the spin-induced quadrupole moment,''
Phys. Rev. D \textbf{104}, 084056 (2021)
[arXiv:2106.09193 [gr-qc]].

\bibitem{Henry:2020ski_v4}
Q.~Henry, G.~Faye and L.~Blanchet,
``Erratum II: Tidal effects in the gravitational-wave phase evolution of compact binary systems to next-to-next-to-leading post-Newtonian order,''
Phys. Rev. D \textbf{111}, 029901(E) (2025)
[version 4 of arXiv:2005.13367 [gr-qc]].

\bibitem{Narikawa_Erratum}
 T.~Narikawa, Erratum Phys. Rev. D, To be published. 

\bibitem{Hinderer:2007mb} 
  T.~Hinderer,
  ``Tidal Love numbers of neutron stars,''
  Astrophys.\ J.\  {\bf 677}, 1216 (2008)
  [arXiv:0711.2420 [astro-ph]].

\bibitem{Flanagan:2007ix} 
  E.~E.~Flanagan and T.~Hinderer,
  ``Constraining neutron star tidal Love numbers with gravitational wave detectors,''
  Phys.\ Rev.\ D {\bf 77}, 021502 (2008)
  [arXiv:0709.1915 [astro-ph]].

\bibitem{Hinderer:2009ca} 
  T.~Hinderer, B.~D.~Lackey, R.~N.~Lang and J.~S.~Read,
  ``Tidal deformability of neutron stars with realistic equations of state and their gravitational wave signatures in binary inspiral,''
  Phys.\ Rev.\ D {\bf 81}, 123016 (2010)
  [arXiv:0911.3535 [astro-ph.HE]].  
  
\bibitem{Vines:2011ud} 
  J.~Vines, E.~E.~Flanagan and T.~Hinderer,
  ``Post-1-Newtonian tidal effects in the gravitational waveform from binary inspirals,''
  Phys.\ Rev.\ D {\bf 83}, 084051 (2011)
  [arXiv:1101.1673 [gr-qc]].

\bibitem{Veitch:2014wba} 
  J.~Veitch {\it et al.},
  ``Parameter estimation for compact binaries with ground-based gravitational-wave observations using the LALInference software library'',
  Phys.\ Rev.\ D {\bf 91}, 042003 (2015)
  [arXiv:1409.7215 [gr-qc]].

\bibitem{LAL} 
 LIGO Scientific Collaboration,
 LIGO Algorithm Library - LALSuite,
 Free Software (GPL), 2018; \texttt{https://doi.org/10.7935/GT1W-FZ16}.
 
\bibitem{Abdelsalhin:2018reg}
T.~Abdelsalhin, L.~Gualtieri and P.~Pani,
``Post-Newtonian spin-tidal couplings for compact binaries,''
Phys. Rev. D \textbf{98}, 104046 (2018)
[arXiv:1805.01487 [gr-qc]].

\bibitem{Banihashemi:2018xfb}
B.~Banihashemi and J.~Vines,
``Gravitomagnetic tidal effects in gravitational waves from neutron star binaries,''
Phys. Rev. D \textbf{101}, 064003 (2020)
[arXiv:1805.07266 [gr-qc]].

\bibitem{JimenezForteza:2018rwr}
X.~Jim\'enez Forteza, T.~Abdelsalhin, P.~Pani and L.~Gualtieri,
``Impact of high-order tidal terms on binary neutron-star waveforms,''
Phys. Rev. D \textbf{98}, 124014 (2018)
[arXiv:1807.08016 [gr-qc]].

\bibitem{Castro:2022mpw}
G.~Castro, L.~Gualtieri, A.~Maselli and P.~Pani,
``Impact and detectability of spin-tidal couplings in neutron star inspirals,''
Phys. Rev. D \textbf{106}, 024011 (2022)
[arXiv:2204.12510 [gr-qc]].

\bibitem{Landry:2018bil}
P.~Landry,
``Rotational-tidal phasing of the binary neutron star waveform,''
[arXiv:1805.01882 [gr-qc]].

\bibitem{Favata:2013rwa} 
  M.~Favata,
  ``Systematic parameter errors in inspiraling neutron star binaries,''
  Phys.\ Rev.\ Lett.\  {\bf 112}, 101101 (2014)
  [arXiv:1310.8288 [gr-qc]].

\bibitem{Wade:2014vqa} 
  L.~Wade, J.~D.~E.~Creighton, E.~Ochsner, B.~D.~Lackey, B.~F.~Farr, T.~B.~Littenberg and V.~Raymond,
  ``Systematic and statistical errors in a bayesian approach to the estimation of the neutron-star equation of state using advanced gravitational wave detectors'',
  Phys.\ Rev.\ D {\bf 89}, 103012 (2014)
  [arXiv:1402.5156 [gr-qc]].
  
\bibitem{Kawaguchi:2018gvj}
K.~Kawaguchi, K.~Kiuchi, K.~Kyutoku, Y.~Sekiguchi, M.~Shibata and K.~Taniguchi,
``Frequency-domain gravitational waveform models for inspiraling binary neutron stars,''
Phys. Rev. D \textbf{97}, 044044 (2018)
[arXiv:1802.06518 [gr-qc]].
    
\bibitem{Dietrich:2017aum} 
  T.~Dietrich, S.~Bernuzzi and W.~Tichy,
  ``Closed-form tidal approximants for binary neutron star gravitational waveforms constructed from high-resolution numerical relativity simulations'',
  Phys.\ Rev.\ D {\bf 96}, 121501(R) (2017)
  [arXiv:1706.02969 [gr-qc]].

\bibitem{Dietrich:2018uni}
T.~Dietrich, S.~Khan, R.~Dudi, S.~J.~Kapadia, P.~Kumar, A.~Nagar, F.~Ohme, F.~Pannarale, A.~Samajdar and S.~Bernuzzi, \textit{et al.}
``Matter imprints in waveform models for neutron star binaries: Tidal and self-spin effects,''
Phys. Rev. D \textbf{99}, 024029 (2019)
[arXiv:1804.02235 [gr-qc]].

\bibitem{Dietrich:2019kaq} 
  T.~Dietrich, A.~Samajdar, S.~Khan, N.~K.~Johnson-McDaniel, R.~Dudi and W.~Tichy,
  ``Improving the NRTidal model for binary neutron star systems,''
  Phys.\ Rev.\ D {\bf 100}, 044003 (2019)
  [arXiv:1905.06011 [gr-qc]].

\bibitem{Yagi:2013sva}
K.~Yagi,
``Multipole Love Relations,''
Phys. Rev. D \textbf{89}, 043011 (2014)
[erratum: Phys. Rev. D \textbf{96}, 129904 (2017); erratum: Phys. Rev. D \textbf{97}, 129901 (2018)]
[arXiv:1311.0872 [gr-qc]].

\bibitem{Dietrich:2020eud}
T.~Dietrich, T.~Hinderer and A.~Samajdar,
``Interpreting Binary Neutron Star Mergers: Describing the Binary Neutron Star Dynamics, Modelling Gravitational Waveforms, and Analyzing Detections,''
Gen. Rel. Grav. \textbf{53}, 27 (2021)
[arXiv:2004.02527 [gr-qc]].

\bibitem{Isoyama:2020lls}
S.~Isoyama, R.~Sturani and H.~Nakano,
``Post-Newtonian templates for gravitational waves from compact binary inspirals,''
in Handbook of Gravitational Wave Astronomy,
edited by C.~Bambi, S.~Katsanevas, and K.~D.~Kokkotas 
(Springer, Singapore, 2021), pp. 1-49
[arXiv:2012.01350 [gr-qc]].

\bibitem{Punturo:2010zz}
M.~Punturo, M.~Abernathy, F.~Acernese, B.~Allen, N.~Andersson, K.~Arun, F.~Barone, B.~Barr, M.~Barsuglia and M.~Beker, \textit{et al.}
``The Einstein Telescope: A third-generation gravitational wave observatory,''
Class. Quant. Grav. \textbf{27}, 194002 (2010).

\bibitem{P1600143}
 Einstein Telecope anticipated sensitivity curve, LIGO Document P1600143-v18, \texttt{https://dcc.ligo.org/LIGO-P1600143/public}.

\bibitem{Branchesi:2023mws}
M.~Branchesi, M.~Maggiore, D.~Alonso, C.~Badger, B.~Banerjee, F.~Beirnaert, E.~Belgacem, S.~Bhagwat, G.~Boileau and S.~Borhanian, \textit{et al.}
``Science with the Einstein Telescope: a comparison of different designs,''
[arXiv:2303.15923 [gr-qc]].

\bibitem{Puecher:2023twf}
A.~Puecher, A.~Samajdar and T.~Dietrich,
``Measuring tidal effects with the Einstein Telescope: A design study,''
[arXiv:2304.05349 [astro-ph.IM]].

\bibitem{PyCBC}
Nitz, A. H., Harry, I. W., Willis, J. L., et al. 
``PyCBC''
Software, \texttt{https://github.com/gwastro/pycbc}, GitHub.

\bibitem{Usman:2015kfa}
S.~A.~Usman, A.~H.~Nitz, I.~W.~Harry, C.~M.~Biwer, D.~A.~Brown, M.~Cabero, C.~D.~Capano, T.~Dal Canton, T.~Dent and S.~Fairhurst, \textit{et al.}
``The PyCBC search for gravitational waves from compact binary coalescence,''
Class. Quant. Grav. \textbf{33}, 215004 (2016)
[arXiv:1508.02357 [gr-qc]].

\bibitem{Dhurandhar:1992mw}
S.~V.~Dhurandhar and B.~S.~Sathyaprakash,
``Choice of filters for the detection of gravitational waves from coalescing binaries. 2. Detection in colored noise,''
Phys. Rev. D \textbf{49}, 1707-1722 (1994)

\bibitem{Buonanno:2009zt} 
  A.~Buonanno, B.~Iyer, E.~Ochsner, Y.~Pan and B.~S.~Sathyaprakash,
  ``Comparison of post-Newtonian templates for compact binary inspiral signals in gravitational-wave detectors'',
  Phys.\ Rev.\ D {\bf 80}, 084043 (2009)
  [arXiv:0907.0700 [gr-qc]].

\bibitem{Blanchet:2013haa} 
  L.~Blanchet,
  Gravitational Radiation from Post-Newtonian Sources and Inspiralling Compact Binaries,
  Living Rev.\ Rel.\  {\bf 17}, 2 (2014)
  [arXiv:1310.1528 [gr-qc]].

\bibitem{Blanchet:2023bwj}
L.~Blanchet, G.~Faye, Q.~Henry, F.~Larrouturou and D.~Trestini,
``Gravitational-Wave Phasing of Compact Binary Systems to the Fourth-and-a-Half post-Newtonian Order,''
[arXiv:2304.11185 [gr-qc]].

\bibitem{Bohe:2013cla} 
  A.~Bohe, S.~Marsat and L.~Blanchet,
  ``Next-to-next-to-leading order spin-orbit effects in the gravitational wave flux and orbital phasing of compact binaries'',
  Class.\ Quant.\ Grav.\  {\bf 30}, 135009 (2013)
  [arXiv:1303.7412 [gr-qc]].

\bibitem{Arun:2008kb} 
  K.~G.~Arun, A.~Buonanno, G.~Faye and E.~Ochsner,
  ``Higher-order spin effects in the amplitude and phase of gravitational waveforms emitted by inspiraling compact binaries: Ready-to-use gravitational waveforms'',
  Phys.\ Rev.\ D {\bf 79}, 104023 (2009)
  Erratum: [Phys.\ Rev.\ D {\bf 84}, 049901 (2011)]
  [arXiv:0810.5336 [gr-qc]].
  
\bibitem{Mikoczi:2005dn} 
  B.~Mikoczi, M.~Vasuth and L.~A.~Gergely,
  ``Self-interaction spin effects in inspiralling compact binaries'',
  Phys.\ Rev.\ D {\bf 71}, 124043 (2005)
  [astro-ph/0504538].

\bibitem{Skilling:2004}
 J.~Skilling, 
 "Nested Sampling,"
 Bayesian Inference and Maximum Entropy Methods in Science and Engineering MAXENT 2004 
 (eds Fischer, R., Dose, V., Preuss, R. \& von Toussaint, U.)
395 (AIP, 2004).

\bibitem{Skilling:2006}
 J.~Skilling, 
 "Nested sampling for general Bayesian computation,"
 Bayesian Analysis {\bf 1}, 833 (2006).
 
\bibitem{Ashton:2022grj}
G.~Ashton, N.~Bernstein, J.~Buchner, X.~Chen, G.~Cs\'anyi, A.~Fowlie, F.~Feroz, M.~Griffiths, W.~Handley and M.~Habeck, \textit{et al.}
``Nested sampling for physical scientists,''
Nature \textbf{2}, 39 (2022)
[arXiv:2205.15570 [stat.CO]].

\bibitem{GWOSC}
\texttt{https://www.gw-openscience.org}.

\bibitem{LIGOScientific:2019lzm}
R.~Abbott \textit{et al.} [LIGO Scientific and Virgo],
``Open data from the first and second observing runs of Advanced LIGO and Advanced Virgo,''
SoftwareX \textbf{13}, 100658 (2021)
[arXiv:1912.11716 [gr-qc]].

\bibitem{Cornish:2014kda}
N.~J.~Cornish and T.~B.~Littenberg,
``BayesWave: Bayesian Inference for Gravitational Wave Bursts and Instrument Glitches,''
Class. Quant. Grav. \textbf{32}, no.13, 135012 (2015)
[arXiv:1410.3835 [gr-qc]].

\bibitem{Littenberg:2015kpb}
T.~B.~Littenberg, J.~B.~Kanner, N.~J.~Cornish and M.~Millhouse,
``Enabling high confidence detections of gravitational-wave bursts,''
Phys. Rev. D \textbf{94}, no.4, 044050 (2016)
[arXiv:1511.08752 [gr-qc]].

\bibitem{Chatziioannou:2019zvs}
K.~Chatziioannou, C.~J.~Haster, T.~B.~Littenberg, W.~M.~Farr, S.~Ghonge, M.~Millhouse, J.~A.~Clark and N.~Cornish,
``Noise spectral estimation methods and their impact on gravitational wave measurement of compact binary mergers,''
Phys. Rev. D \textbf{100}, no.10, 104004 (2019)
[arXiv:1907.06540 [gr-qc]].

\bibitem{Soares-Santos:2017lru} 
  DES and Dark Energy Camera GW-EM Collaborations: M.~Soares-Santos, D.~E.~Holz, J.~Annis, R.~Chornock, K.~Herner, E.~Berger, D.~Brout, H.~Chen, R.~Kessler, M.~Sako {\it et al.},
  ``The Electromagnetic Counterpart of the Binary Neutron Star Merger LIGO/Virgo GW170817. I. Discovery of the Optical Counterpart Using the Dark Energy Camera,''
  Astrophys.\ J.\ Lett. {\bf 848}, L16 (2017)
  [arXiv:1710.05459 [astro-ph.HE]].

\bibitem{LIGOScientific:2017ync}
B.~P.~Abbott \textit{et al.} [LIGO Scientific, Virgo, Fermi GBM, INTEGRAL, IceCube, AstroSat Cadmium Zinc Telluride Imager Team, IPN, Insight-Hxmt, ANTARES, Swift, AGILE Team, 1M2H Team, Dark Energy Camera GW-EM, DES, DLT40, GRAWITA, Fermi-LAT, ATCA, ASKAP, Las Cumbres Observatory Group, OzGrav, DWF (Deeper Wider Faster Program), AST3, CAASTRO, VINROUGE, MASTER, J-GEM, GROWTH, JAGWAR, CaltechNRAO, TTU-NRAO, NuSTAR, Pan-STARRS, MAXI Team, TZAC Consortium, KU, Nordic Optical Telescope, ePESSTO, GROND, Texas Tech University, SALT Group, TOROS, BOOTES, MWA, CALET, IKI-GW Follow-up, H.E.S.S., LOFAR, LWA, HAWC, Pierre Auger, ALMA, Euro VLBI Team, Pi of Sky, Chandra Team at McGill University, DFN, ATLAS Telescopes, High Time Resolution Universe Survey, RIMAS, RATIR and SKA South Africa/MeerKAT],
``Multi-messenger Observations of a Binary Neutron Star Merger,''
Astrophys. J. Lett. \textbf{848}, L12 (2017)
[arXiv:1710.05833 [astro-ph.HE]].

\bibitem{J-GEM:2017tyx}
Y.~Utsumi \textit{et al.} [J-GEM],
``J-GEM observations of an electromagnetic counterpart to the neutron star merger GW170817,''
Publ. Astron. Soc. Jap. \textbf{69}, 101 (2017)
[arXiv:1710.05848 [astro-ph.HE]].

\bibitem{KAGRA:2013rdx}
B.~P.~Abbott \textit{et al.} [LIGO Scientific, Virgo and KAGRA],
``Prospects for observing and localizing gravitational-wave transients with Advanced LIGO, Advanced Virgo and KAGRA,''
Living Rev.Rel. \textbf{23}, 3 (2020),
Living Rev. Rel. \textbf{21}, 3 (2018)
[arXiv:1304.0670 [gr-qc]].

\bibitem{DelPozzo:2013ala}
W.~Del Pozzo, T.~G.~F.~Li, M.~Agathos, C.~Van Den Broeck and S.~Vitale,
``Demonstrating the feasibility of probing the neutron star equation of state with second-generation gravitational wave detectors,''
Phys. Rev. Lett. \textbf{111}, 071101 (2013)
[arXiv:1307.8338 [gr-qc]].

\bibitem{Lackey:2014fwa}
B.~D.~Lackey and L.~Wade,
``Reconstructing the neutron-star equation of state with gravitational-wave detectors from a realistic population of inspiralling binary neutron stars,''
Phys. Rev. D \textbf{91}, 043002 (2015)
[arXiv:1410.8866 [gr-qc]].

\bibitem{Wysocki:2020myz}
D.~Wysocki, R.~O'Shaughnessy, L.~Wade and J.~Lange,
``Inferring the neutron star equation of state simultaneously with the population of merging neutron stars,''
[arXiv:2001.01747 [gr-qc]].

\bibitem{Dudi:2018jzn}
R.~Dudi, F.~Pannarale, T.~Dietrich, M.~Hannam, S.~Bernuzzi, F.~Ohme and B.~Br\"ugmann,
``Relevance of tidal effects and post-merger dynamics for binary neutron star parameter estimation,''
Phys. Rev. D \textbf{98}, 084061 (2018)
[arXiv:1808.09749 [gr-qc]].

\bibitem{Samajdar:2018dcx} 
  A.~Samajdar and T.~Dietrich,
  ``Waveform systematics for binary neutron star gravitational wave signals: effects of the point-particle baseline and tidal descriptions,''
  Phys.\ Rev.\ D {\bf 98}, 124030 (2018)
  [arXiv:1810.03936 [gr-qc]].

\bibitem{Messina:2019uby}
F.~Messina, R.~Dudi, A.~Nagar and S.~Bernuzzi,
``Quasi-5.5PN TaylorF2 approximant for compact binaries: point-mass phasing and impact on the tidal polarizability inference,''
Phys. Rev. D \textbf{99}, 124051 (2019)
[arXiv:1904.09558 [gr-qc]].

\bibitem{Samajdar:2019ulq}
A.~Samajdar and T.~Dietrich,
``Waveform systematics for binary neutron star gravitational wave signals: Effects of spin, precession, and the observation of electromagnetic counterparts,''
Phys. Rev. D \textbf{100}, 024046 (2019)
[arXiv:1905.03118 [gr-qc]].

\bibitem{Agathos:2019sah}
M.~Agathos, F.~Zappa, S.~Bernuzzi, A.~Perego, M.~Breschi and D.~Radice,
``Inferring Prompt Black-Hole Formation in Neutron Star Mergers from Gravitational-Wave Data,''
Phys. Rev. D \textbf{101}, 044006 (2020)
[arXiv:1908.05442 [gr-qc]].

\bibitem{Landry:2020vaw}
P.~Landry, R.~Essick and K.~Chatziioannou,
``Nonparametric constraints on neutron star matter with existing and upcoming gravitational wave and pulsar observations,''
Phys. Rev. D \textbf{101}, 123007 (2020)
[arXiv:2003.04880 [astro-ph.HE]].

\bibitem{Chen:2020fzm}
A.~Chen, N.~K.~Johnson-McDaniel, T.~Dietrich and R.~Dudi,
``Distinguishing high-mass binary neutron stars from binary black holes with second- and third-generation gravitational wave observatories,''
Phys. Rev. D \textbf{101}, 103008 (2020)
[arXiv:2001.11470 [astro-ph.HE]].

\bibitem{Chatziioannou:2021tdi}
K.~Chatziioannou,
``Uncertainty limits on neutron star radius measurements with gravitational waves,''
Phys. Rev. D \textbf{105}, 084021 (2022)
[arXiv:2108.12368 [gr-qc]].

\bibitem{Kunert:2021hgm}
N.~Kunert, P.~T.~H.~Pang, I.~Tews, M.~W.~Coughlin and T.~Dietrich,
``Quantifying modelling uncertainties when combining multiple gravitational-wave detections from binary neutron star sources,''
Phys. Rev. D \texttt{105}, L061301 (2022)
[arXiv:2110.11835 [astro-ph.HE]].


\end{thebibliography}
\end{document}